\documentclass[aps,prc,twocolumn,groupedaddress]{revtex4}

\usepackage{amsmath,amssymb}
\usepackage[dvipdfm]{graphicx,color}
\usepackage{subfigure}
\usepackage{longtable}

\begin{document}


\title{Dineutron formation and breaking in $^8$He}


\author{Fumiharu Kobayashi and Yoshiko Kanada-En'yo}
\affiliation{Department of Physics, Kyoto University, Kyoto 606-8502, Japan}


\date{\today}

\begin{abstract}
We discuss the correlation of the spin-singlet two-neutron pair, the dineutron correlation, 
in the ground and excited $0^+$ states of $^8$He
by using a method to describe the dineutron correlation in neutron-rich nuclei, 
the dineutron condensate wave function.
The shell-model configuration 
and the dineutron configuration compete with each other in the ground state of $^8$He.
We compare the ground state structure of $^8$He with that of $^6$He
and suggest that the dineutron correlation is weaker
in the ground state of $^8$He than $^6$He. 
We also suggest the possibility of the dineutron condensation in an excited $0^+$ state of $^8$He. 
In addition, we investigate the formation mechanism 
of the dineutron condensation in $^8$He
by using a newly developed framework, $\alpha$ and dineutron condensate wave function, 
and conclude that the attraction from the core is essential
for the formation of the dineutron condensation.
\end{abstract}

\pacs{}

\maketitle

\section{Introduction}
\label{secI}

In unstable nuclei, a lot of exotic phenomena have been discovered
and they are eagerly being investigated both theoretically and experimentally.
The dineutron correlation is one of the interesting topics of neutron-rich nuclei. 
A dineutron is a spin-singlet pair of two neutrons. 
As is known, the neutron-neutron interaction is attractive in the spin-singlet channel, 
but two neutrons are not bound in free space. 
Nevertheless, it is suggested that two neutrons form a spatially compact dineutron 
in some situations such as in the low-density region of the nuclear matter 
\cite{baldo90, matsuo06}
or in the neutron-halo and -skin regions of neutron-rich nuclei
\cite{bertsch91, zhukov93, matsuo05, hagino05, hagino07}. 
The spatial correlation between two neutrons in the infinite nuclear matter and in the finite nuclei
has been discussed in association with the BCS-BEC crossover. 
Although a compact dineutron can be formed especially in finite nuclei, 
a dineutron is naturally so soft and fragile 
that its size or amount of its component would change 
very readily depending on the circumstances such as the nuclear density. 
However, the properties of the dineutron correlation has not been well-known yet, 
so our aim is to clarify the formation mechanism 
and the universal properties of the dineutron correlation. 
Once a compact dineutron is formed, it might behave as a kind of boson 
or a kind of cluster in finite nuclei. 
The understanding of the dineutron formation mechanism 
will also lead to the more sophisticated understanding of the general cluster formation mechanism. 
In order to investigate the dineutron correlation in finite nuclei, 
we constructed the method of the dineutron condensate (DC) wave function \cite{kobayashi11}, 
which is useful to investigate dineutron behavior in various nuclei. 
In this work, we apply it to $^8$He 
and investigate the structures of $0^+$ states from the viewpoint of the dineutron correlation. 
$^8$He has four valence neutrons so that 
not only one dineutron but two dineutrons can be formed. 
It is challenging to investigate the mechanism of the development of one or two dineutrons 
and to make clear how these dineutrons behave 
in the ground and excited states in $^8$He. 

The dineutron correlation is often discussed in relation with the $\alpha$ condensation, 
which has been suggested in dilute nuclear matter \cite{roepke98}. 
The idea of the $\alpha$ condensation was also applied 
to understand developed cluster structures in excited states of stable nuclei. 
For example, it was suggested that
$^{12}$C$(0^+_2)$ known as the representative $\alpha$ cluster state 
has an exotic gas-like structure
where three $\alpha$ clusters are in the $S$-orbit and weakly interacting each other, 
and this state is interpreted as an $\alpha$ condensation
\cite{tohsaki01, funaki03, funaki05, funaki09}. 
The idea of the $\alpha$ condensation is also extended to the nuclei 
other than $^{12}$C of three $\alpha$s, 
for example $^{16}$O of four $\alpha$s \cite{tohsaki01, funaki08}
and $^{24}$Mg of two $\alpha$s around an $^{16}$O core \cite{itagaki07, ichikawa11}.  
In analogy with the $\alpha$ condensate states, 
the dineutron condensate states in $^8$He and $^7$H were also discussed
in the theoretical works studying the dineutron correlation in He and H isotopes
\cite{enyo07, itagaki08, aoyama09}. 
Those studies suggested the possibility of 
the dineutron condensate states in $^8$He or $^7$H, 
but still those are not sufficient to make clear the formation mechanism
of the dineutron condensation. 
We would like to answer the unsolved questions; what kind of condition enhances the dineutron correlation
and how is the dineutron motion activated to form the dineutron condensation?

The aims of the present work are, first, 
to make clear how the one- or two- dineutron components 
contribute to the ground and excited states of $^8$He 
and how much such dineutron components and the shell-model components are mixed. 
Moreover, we compare the structure of the ground state of $^8$He and that of $^6$He
and discuss the analogy and difference of the dineutron correlation in them. 
Second, we investigate the possibility of a two-dineutron condensate state 
in an excited $0^+$ state of $^8$He,  
and then clarify the formation mechanism of the dineutron condensation
with an analysis of an ideal $\alpha+2n+2n$ three-body condensate state. 

The contents of this paper are the following.  
We investigate the dineutron correlation in  $^8$He in Sec.~\ref{secII}. 
There we discuss about the contribution of the dineutron correlation to the ground state, 
and compare it with that in the ground state of $^6$He.
In addition, we suggest the excited dineutron condensate state of $^8$He.
In Sec.~\ref{secIII}, we give a discussion of 
the formation mechanism of the dineutron condensation in $^8$He
by analyzing the ideal $\alpha+2n+2n$ system. 
We examine the energy behavior of two dineutrons around the $\alpha$ core 
and compare the result with that around a triton core
to show that the potential from the core plays an important role for the dineutron condensation. 
Finally, we summarize the present work in Sec.~\ref{secIV}.

\section{Dineutron correlation in $^8$He}
\label{secII}

In this section, we mainly investigate the dineutron correlation in $^8$He. 
We discuss the dineutron component in the ground state of $^8$He, 
and compare it with that in the ground state of $^6$He. 
We also suggest
the possibility of the dineutron condensate state in an excited $0^+$ state of $^8$He. 

\subsection{Framework}
\label{secII-A}
Here we describe the framework for the study of $^{6,8}$He. 
We first explain the method applied to $^8$He, 
and then we give the explanation of the one for $^6$He. 

We suppose that not only the $0p$-shell neutron configurations but also 
the strongly correlated two-neutron pairs, 
that is the cluster-like configurations of developed dineutrons, would be important
in light neutron-rich nuclei. 
In this work, we are interested in the behavior of valence neutrons around an $\alpha$ core, 
especially in the formation and breaking of the dineutrons. 
Therefore, we quantify the size and the center of mass motion of the dineutron
as well as the degree of dissociation. 
In order to consider the behavior of the valence neutrons fully, 
we superpose two kinds of wave function to describe $^8$He, 
the extended $^6$He$+2n$ cluster wave functions 
and $^8$He DC wave functions. 
The extended $^6$He$+2n$ cluster wave functions mainly describe
the configurations containing a dineutron around $^6$He 
and also shell-model configurations of four neutrons around an $\alpha$ core. 
On the other hand, $^8$He DC wave function
can describe the ones containing one- or two-dineutron developed spatially.

\subsubsection{Extended $^6$He$+2n$ cluster wave function}
\label{secII-A-1}
The present $^6$He$+2n$ cluster wave function is 
an extended one where the dissociation of the $2n$ cluster is taken into account. 
First, we explain the treatment of the extended $2n$ cluster, 
denoted as the $2n^*$ cluster hereafter. 
In the cluster model, the $2n$ cluster wave function, $\Phi_{2n}$, is composed of two neutrons 
described with the Gaussian wave packet. 
\begin{align}
\Phi_{2n}(\boldsymbol{R}, b) = 
\mathcal{A} \{ & \phi_n (\boldsymbol{r}_1; \boldsymbol{R}, b)
\chi_{\uparrow}(1) \nonumber \\
& \times \phi_n (\boldsymbol{r}_2; \boldsymbol{R}, b)
\chi_{\downarrow}(2) \}, \label{eq:2n_wf} \\
\phi_n (\boldsymbol{r};  \boldsymbol{R}, b)
\propto &\ \exp \left[ - \frac{1}{2b^2} 
\left( \boldsymbol{r} - \boldsymbol{R} \right)^2 \right], 
\label{eq:Gaussian_wf}
\end{align}
where $\chi_{\uparrow, \downarrow}(i) \ (i = 1,2)$ is
the spin-up or -down wave function of the $i$th neutron. 
This wave function contains two Gaussian parameters, the center parameter, $\boldsymbol{R}$, 
and the size parameter, $b$.
In the ordinary cluster wave functions, 
the Gaussian center, $\boldsymbol{R}$, is a real value, 
which corresponds to 
the mean position of the single particle, $\langle \boldsymbol{r} \rangle$. 
When it is a complex value,  
the imaginary part corresponds to its mean momentum, $\langle \boldsymbol{p} \rangle$. 
\begin{align}
\langle \boldsymbol{r} \rangle = &\ {\rm Re} [\boldsymbol{R}], \\
\langle \boldsymbol{p} \rangle = &\ \frac{\hbar}{b^2}  {\rm Im} [\boldsymbol{R}].
\end{align}
In the extended $2n$ cluster wave function, $\Phi_{2n^*}$, 
we use the complex values for the Gaussian centers. 
In this work, we define the complex Gaussian center parameters 
for the spin-up and -down neutrons in the $2n^*$ cluster
as $\boldsymbol{R}_{\uparrow/\downarrow} = \boldsymbol{R}_{\lambda}
\pm i \lambda \boldsymbol{e}_{\lambda}$.
We fix the real part for two neutrons
to the same value $\boldsymbol{R}_{\lambda}$ to locate them at the same position. 
We introduce an imaginary part proportional to the real factor $\lambda$, 
which characterizes the degree of dissociation of the $2n^*$ cluster, 
in the $\boldsymbol{e}_{\lambda}$-direction. 
$\boldsymbol{e}_{\lambda}$ is orthogonalized to 
both the direction of $\hat{\boldsymbol{R}}_{\lambda}$
(the unit vector of $\boldsymbol{R}_{\lambda}$)
and that of the spin-quantized axis $\boldsymbol{e}_{\rm spin}$,  
as $\boldsymbol{e}_{\lambda} \equiv \boldsymbol{e}_{\rm spin}
\times \hat{\boldsymbol{R}}_{\lambda}$. 
Here we choose $\boldsymbol{e}_z$ as $\boldsymbol{e}_{\rm spin}$, 
$\boldsymbol{e}_x$ as $\hat{\boldsymbol{R}}_{\lambda}$
and $\boldsymbol{e}_y$ as $\boldsymbol{e}_{\lambda}$. 
$\boldsymbol{R}_{\uparrow, \downarrow}$ has
the imaginary part with the opposite sign for each spin-up and -down neutron, 
so we give the opposite momentum to each neutron
in order to take into account the dissociation of the $2n^*$ cluster 
due to the spin-orbit interaction,
as is done for an $\alpha$ cluster in Refs.~\cite{itagaki05, masui07}. 
Then the $2n^*$ cluster wave function, $\Phi_{2n^*}$, can be written as
\begin{align}
\Phi_{2n^*}(\lambda, \boldsymbol{R}_{\lambda}, b_{\lambda}) = 
\mathcal{A} \{ & \phi_n (\boldsymbol{r}_1; \boldsymbol{R}_{\uparrow}, b_{\lambda})
\chi_{\uparrow}(1) \nonumber \\
& \times \phi_n (\boldsymbol{r}_2; \boldsymbol{R}_{\downarrow}, b_{\lambda})
\chi_{\downarrow}(2) \}.
\label{eq:2n*_wf}
\end{align}
In addition to two parameters in Eq.~(\ref{eq:2n_wf}), 
that is the location, $\boldsymbol{R}_\lambda$, 
and the size, $b_{\lambda}$, of the ordinary $2n$ cluster, 
the $2n^*$ cluster wave function contains 
an additional parameter $\lambda$ 
which corresponds to the degree of breaking of the $2n^*$ cluster. 
These three parameters characterize the behavior of the $2n^*$ cluster
in this extended cluster wave function. 

We superpose the ones with various parameter sets
$\{ \lambda, \boldsymbol{R}_{\lambda}, b_{\lambda} \}$, 
to describe various configurations of two neutrons around the $^6$He core. 
The $^6$He core is assumed to be 
the mixing of the harmonic oscillator (H.O.) $0p$-shell configurations. 
The $^6$He wave function located at $\boldsymbol{R}_{^6{\rm He}}$ is denoted as 
$\Phi_{^6{\rm He}}(\kappa, \boldsymbol{R}_{^6{\rm He}}, b_{\alpha})$, 
where $\kappa$ assigns the $(0p)^2$-configuration of two neutrons around the $\alpha$, 
and the size parameters of the single particle wave functions in the $^6$He core 
are fixed to the same value as $b_{\alpha} = 1.46$ fm. 
Practically, we describe the $^6$He core with the shifted Gaussians 
as is done in Refs.~\cite{enyo12,kobayashi12}. 

The extended $^6$He$+2n$ 
($^6$He$+2n^*$) cluster wave function is composed of 
the $^6$He and $2n^*$ wave functions mentioned above. 
In this framework, we locate the mean position of the total center of mass at the origin, 
$(6\boldsymbol{R}_{^6{\rm He}} + 2\boldsymbol{R}_{\lambda})/8 = 0$, 
and choose $b_{\lambda} = b_{\alpha}$
to extract the center of mass motion exactly.   
The $^6$He$+2n^*$ cluster wave function can be written down as follows. 
\begin{align}
& \Phi_{^6 {\rm He}+2n^*} (\kappa; \lambda, d_{\lambda}, b_{\lambda}) \nonumber \\
&\ = \frac{1}{\sqrt{8!}} \mathcal{A} \{ 
\Phi_{^6 {\rm He}} (\kappa, \boldsymbol{R}_{^6{\rm He}} = 
- \frac{1}{4}d_{\lambda} \boldsymbol{e}_x, b_{\alpha}) \nonumber \\
& \hspace{3.5em} \times
\Phi_{2n^*}(\lambda, \boldsymbol{R}_{\lambda} = 
\frac{3}{4}d_{\lambda} \boldsymbol{e}_x, b_{\lambda} = b_{\alpha}) \}.
\end{align}
In this work, we choose $\lambda = 0.0$ or 0.4, 
which corresponds to how much the $2n^*$ cluster is broken, 
and $d_{\lambda} = 1, 2, 3, 4 $ fm for each $\lambda$. 
In the case of $\lambda = 0.0$, 
two neutrons form a spin-singlet cluster, that is a dineutron, around the $^6$He core. 
On the other hand, when $\lambda = 0.4$, 
spin-up and -down neutrons have the opposite momentum each other 
and the $2n^*$ cluster is dissociated due to the spin-orbit potential from the core. 
In the $d_{\lambda} \rightarrow 0$ limit, 
linear combination of these two $\lambda$ values describes 
two neutrons in the $(0p)^2$-orbits including the $(0p_{3/2})^2$ configuration. 
Therefore, by superposing these two kinds of $\lambda$ value, 
in one limit, the $^6$He$+2n^*$ cluster wave function does describe 
the $(0p_{3/2})^4$ sub-shell closed configuration, 
and in another limit, it does describe 
the $^6$He core with $0p$-shell configurations 
and two neutrons forming a dineutron around the $0p$-shell $^6$He core.

\subsubsection{$^8$He DC wave function}
\label{secII-A-2}
In addition to the $^6$He$+2n^*$ cluster wave functions explained above, 
we superpose the $^8$He DC wave functions 
to describe states containing largely developed one or two dineutrons. 
In the DC wave function, the system is composed of a core and one or a few dineutrons, 
which is an $\alpha+2n^*$ core and one dineutron in the present $^8$He case. 
The explicit dineutron is distributed around the core in the $S$-wave.
(When we mention the orbit of the cluster, we use the capital letter such as $S$.)
The details of the DC wave function are referred in Ref.~\cite{kobayashi11}. 
The $^8$He DC wave function used in this work 
is composed of the $\alpha$ and $2n^*$ clusters and one explicit dineutron as 
\begin{align}
& \Phi_{\rm DC} (\lambda, d_{\lambda}, b_{\lambda}; B_n, b_n) \nonumber \\ 
&\ = \frac{1}{\sqrt{8!}} \int d^3 \boldsymbol{R} \exp \left[ - \frac{R^2}{B_n^2} \right]
\nonumber \\
&\ \hspace{2em} \times 
\mathcal{A} \{ \Phi_{\alpha} 
(\boldsymbol{R}_{\alpha} = - \frac{1}{3}d_{\lambda} \boldsymbol{e}_x, b_{\alpha}) 
\nonumber \\
&\ \hspace{2em} \times 
\Phi_{2n^*} 
(\lambda, \boldsymbol{R}_{\lambda} = \frac{2}{3}d_{\lambda} \boldsymbol{e}_x, 
b_{\lambda} = b_{\alpha}) \
\Phi_{2n} (\boldsymbol{R}, b_n) \}. 
\label{eq:DC_wf}
\end{align}
$\Phi_{\alpha}$ is the $\alpha$ cluster wave function
composed of a set of $(p\uparrow, p\downarrow, n\uparrow, n\downarrow)$ 
located at $\boldsymbol{R}_{\alpha}$ 
and its size parameter is $b_{\alpha} = 1.46$ fm. 
$\Phi_{2n, 2n^*}$ is the ordinary and extended $2n$ cluster wave function as Eq.~(\ref{eq:2n_wf}) and (\ref{eq:2n*_wf}). 
The ordinary $2n$ cluster in the DC wave function
is the explicit dineutron in the $S$-wave around the core
by performing the integral of the Gaussian center, $\boldsymbol{R}$. 
Its wave function, $\Phi_{2n_{\rm DC}}$, 
is written after the $\boldsymbol{R}$-integral in Eq.~(\ref{eq:DC_wf}) as follows. 
\begin{align}
\Phi_{2n_{\rm DC}} (\beta, b_n) = &\ \int d^3 \boldsymbol{R} \exp \Big[ - \frac{R^2}{B_n^2} \Big]
\Phi_{2n}(\boldsymbol{R},b_n) \nonumber \\
= &\ \mathcal{A} \{ \psi_r (b_n) \psi_G(\beta) \ 
\chi_{\uparrow}(n1) \chi_{\downarrow}(n2) \}, \\
\psi_r (\boldsymbol{r};b_n) & \propto
\exp \left[ - \frac{r^2}{4b_n^2} \right] 
\hspace{1em}
\boldsymbol{r} = \boldsymbol{r}_{n1} - \boldsymbol{r}_{n2}, \label{eq:DCwf_rel} \\
\psi_G (\boldsymbol{r}_G;\beta) & \propto 
\exp \left[ - \frac{r_G^2}{\beta^2} \right] 
\hspace{1em}
\boldsymbol{r}_G = \frac{\boldsymbol{r}_{n1} + \boldsymbol{r}_{n2}}{2}, 
\label{eq:DCwf_com}
\end{align}
where we introduce the new parameter $\beta$ defined as $\beta^2 = B_n^2 + b_n^2$
instead of $B_n$. 
$\psi_{r}$ and $\psi_G$ are the relative 
and center of mass wave functions of the dineutron, respectively. 
$\boldsymbol{r}_{n1,2}$ are the coordinates of two neutrons in the explicit dineutron. 
As seen in Eq.~(\ref{eq:DCwf_rel}) and (\ref{eq:DCwf_com}), 
in a DC wave function, 
the relative motion between two neutrons in the dineutron is the $s$-wave, 
and the one between the dineutron and the core is also the $S$-wave. 
The DC wave function contains characteristic parameters $b_n$ and $\beta$. 
$b_n$ is the Gaussian width in the dineutron relative wave function (Eq.~(\ref{eq:DCwf_rel}))
so that this parameter corresponds to the relative distance between two neutrons
in the dineutron, that is the dineutron size.
On the other hand, $\beta$ is 
the width in the dineutron center of mass wave function  (Eq.~(\ref{eq:DCwf_com})), 
and it corresponds to the spatial expansion 
of the dineutron center of mass motion from the core. 
By superposing the DC wave functions with various values of $b_n$ and $\beta$, 
we sufficiently consider the change of the dineutron size 
and the one of the dineutron spatial expansion from the core.

We describe the system containing one- or two-dineutron
by using the DC wave function mentioned above. 
A pair of two neutrons, $\Phi_{2n_{\rm DC}}$, is assumed to be spin-singlet, that is a dineutron cluster, 
and described with the relative and center of mass wave function defined 
in Eq.~(\ref{eq:DCwf_rel}) and (\ref{eq:DCwf_com}). 
The behavior of these neutrons is characterized by 
the parameters of $\beta$ (the dineutron expansion from the core)
and $b_n$ (the dineutron size). 
In this work, we superpose $\beta = 2, 3, \ldots, 8$ fm and five $b_n$ values 
for each $\beta$ chosen as
\begin{equation}
b_i = 1.5 \times \left( \frac{\beta-0.2}{1.5} \right)^{(i-1)/(5-1)} \ ( i = 1, \ldots, 5).
\label{eq:b_n}
\end{equation} 
Since we superpose many $\beta$ and $b_n$ values, 
such a dineutron changes the size largely and it can be distributed far from the core. 
Another pair of two neutrons, $\Phi_{2n^*}$, around the $\alpha$ core is described 
similarly to that in the $^6$He$+2n^*$ cluster wave function. 
The behavior of these neutrons is characterized by
the parameters of $\lambda$ (the degree of $2n^*$ cluster dissociation), 
$d_{\lambda}$ (the distance between $\alpha$ and $2n^*$)
and $b_{\lambda}$ (the size of $2n^*$). 
As in the $^6$He$+2n^*$ cluster wave function, 
we choose $\lambda = 0.0$ or 0.4 to take into account 
the breaking effect of one dineutron in the DC wave function. 
In the $\lambda = 0.0$ bases, 
we choose $d_{\lambda} = 1, 2, \ldots, 8$ fm and $b_{\lambda} = b_n$, 
and then the system contains two developed dineutrons with the same size. 
On the other hand, in the $\lambda = 0.4$ bases, 
we choose $d_{\lambda} = 1, 2, 3, 4$ fm and $b_{\lambda} = b_{\alpha}$
to describe the system containing a dissociated $2n^*$ cluster around the $\alpha$ core
and one developed dineutron in the $S$-wave.
Therefore, by superposing the DC wave functions mentioned above, 
we can describe the $(0p)^2$-configurations of $^6$He
plus a dineutron developed from the $^6$He core in the $d_{\lambda} \rightarrow 0$ limit, 
and also the ones of two dineutrons expanded from the $\alpha$ core
including the two-dineutron condensation 
which means that two dineutrons with the same size
are equally in the same $S$-wave around the $\alpha$ core. 

\begin{table*}[t!]
\caption{The superposed parameters in the $^6$He$+2n^*$ wave functions
and the $^8$He DC wave functions to describe $^8$He$(0^+)$. 
The label ``$2n^*$'' corresponds to the parameters
of the $2n^*$ cluster in each kind of wave function, 
and ``$2n_{\rm DC}$'' to the ones of the explicit dineutron cluster in the DC wave function. }
\label{tab:parameter_He8}
\begin{ruledtabular}
\begin{tabular}{c|lll|rr}
& \multicolumn{3}{c|}{$2n^*$} 
& \multicolumn{2}{c}{$2n_{\rm DC}$} \\ \hline
$\Phi_{^6{\rm He}+2n^*}$ & $\lambda = 0.0, 0.4$ & 
$d_{\lambda} = 1, 2, 3, 4$ & $b_{\lambda} = b_{\alpha}$ & & \\ \hline
$\Phi_{\rm DC}$ & $\lambda=0.0$ & 
$d_{\lambda} = 1, 2, \ldots, 8 $ & $b_{\lambda} = b_n$ & 
$\beta = 2, 3, \ldots, 8$ & $b_n = b_i \ (i = 1, \ldots, 5)$ \\ \cline{2-6}
&  $\lambda=0.4$ & 
$d_{\lambda} = 1, 2, 3, 4$ & $b_{\lambda} = b_{\alpha}$ &
$\beta = 2, 3, \ldots, 8$ & $b_n = b_i \ (i = 1, \ldots, 5)$ \\
\end{tabular}
\end{ruledtabular}
\end{table*}

\subsubsection{Description of $^8$He$(0^+)$}
\label{secII-A-3}
$^8$He$(0^+)$ is described with the linear combination of those wave functions. 
\begin{align}
\Psi_{^8 {\rm He} (0^+)} = &\ \sum_i
c_i \ \mathcal{P}^{0+}_{00} \Phi_{^6 {\rm He}+2n^*} 
(\kappa; \lambda, d_{\lambda}, b_{\lambda} = b_{\alpha}) \nonumber \\
&\ + \sum_j
c_j \ \mathcal{P}^{0+}_{00} 
\Phi_{\rm DC} ( \lambda, d_{\lambda}, b_{\lambda}; \beta, b_n), 
\label{eq:He8_wf}
\end{align}
where $i$ and $j$ are the abbreviations of
$i = \{ \kappa; \lambda, d_{\lambda} \}$ 
and $j =  \{ \lambda, d_{\lambda}, b_{\lambda}; \beta, b_n \}$. 
$\mathcal{P}^{J \pi}_{MK}$ is the projection operator to 
the eigenstate having the total spin-parity of $J^{\pi}$. 
In this work, we discuss only the $0^+$ states in $^8$He. 
The coefficients, $c_{i,j}$, are determined by 
the diagonalization of the Hamiltonian shown in Sec.~\ref{secII-B}.
As mentioned above, 
we superpose many wave functions with various parameters, 
and we show the list of the superposed parameters in Table~\ref{tab:parameter_He8}. 
Here five values of $b_i$ ($i=1, \ldots, 5$) are chosen for each $\beta$ value 
using Eq.~(\ref{eq:b_n}).

Here we repeat the superposed parameters and 
what components of the valence neutrons are described in the present framework. 
For the $2n^*$ cluster around the $^6$He cluster
in the $^6$He$+2n^*$ cluster wave function 
and that around the $\alpha$ cluster in the $^8$He DC wave function, 
we choose $\lambda = 0.0$ or 0.4 for the degree of breaking of the $2n^*$ cluster. 
$\lambda = 0.0$ corresponds to the situation that two neutrons form a dineutron cluster,
and $\lambda = 0.4$ to the one that two neutrons have the opposite momentum
to be dissociated due to the spin-orbit force from the core. 
Then, in the $^6$He$+2n^*$ cluster wave function, 
we choose as $d_{\lambda} = 1, 2, 3, 4$ fm 
and $b_{\lambda} = b_{\alpha} = 1.46$ fm for each $\lambda$, 
which means the $2n^*$ cluster is distributed around the $^6$He core
with the same size parameter as the core. 
By superposing such $^6$He$+2n^*$ cluster wave functions, 
we can describe any $(0p)^4$ neutron configuration
including the $(0p_{3/2})^4$ sub-shell closed configuration
in the small $d_{\lambda}$ limit, 
and also express one dineutron around the $^6$He cluster. 
In the $^8$He DC wave function, 
we choose $d_{\lambda} = 1, 2, \ldots, 8$ fm and $b_{\lambda} = b_n$
for $\lambda = 0.0$. 
In this case, 
two pairs of two neutrons form two spin-singlet dineutron clusters, 
whose sizes change but are the same each other,  
developed largely from the $\alpha$ core. 
In the $\lambda = 0.4$ case, 
we choose $d_{\lambda} = 1, 2, 3, 4$ fm and $b_{\lambda} = b_{\alpha}$, 
and these wave functions describe the system
where a pair of two neutrons is a dissociated $2n^*$ cluster around an $\alpha$ core
and another pair of two neutrons forms a dineutron expanded far from $\alpha+2n^*$. 
Therefore, the model space in the present framework contains 
various configurations from $(0p_{3/2})^4$ sub-shell closure one
to those with two largely developed dineutrons. 
In the next section, we investigate how these components contribute to $^8$He in detail.

\subsubsection{Description of $^6$He$(0^+)$}
\label{secII-A-4}
In the present work, we also investigate the dineutron correlation in $^6$He 
and compare it with that in $^8$He. 
For this aim, we describe the $^6$He wave function
with a similar framework to that for $^8$He. 
That is the superposition of the extended $\alpha+2n$ ($\alpha+2n^*$) cluster wave functions 
and the $^6$He DC wave functions. 
In the $\alpha+2n^*$ cluster wave function, 
the $^6$He cluster in the $^6$He$+2n^*$ one is replaced by an $\alpha$ cluster, 
and the rest is the same.
Also in the $^6$He DC wave function, the core is an $\alpha$. 
Superposing these wave functions, we describe the $0^+$ state of $^6$He.
\begin{align}
\Psi_{^6 {\rm He} (0^+)} = &\ \sum_i
c_i \ \mathcal{P}^{0+}_{00} 
\Phi_{\alpha+2n^*} (\lambda, d_{\lambda}, b_{\lambda} = b_{\alpha}) \nonumber \\
&\ + \sum_j
c_j \ \mathcal{P}^{0+}_{00} \Phi_{\rm DC} (\beta, b_n), 
\label{eq:He6_wf}
\end{align}
where $i$ and $j$ are the abbreviations of
$i = \{ \lambda, d_{\lambda} \}$ 
and $j =  \{ \beta, b_n \}$. 
In the first term of Eq.~(\ref{eq:He6_wf}), 
we superpose $\lambda =  0.0, 0.4$ as done in the description of $^8$He
(Sec.~\ref{secII-A-3}), 
and $d_{\lambda} = 1, 2, 3, 4$ fm for each $\lambda$
so that it can describe the valence neutrons in $(0p)^2$ orbits 
and the ones forming a compact dineutron near the $\alpha$,
that is the same configurations as that in $^8$He, 
if the $^6$He ($(0p)^2$) core is replaced with the $\alpha$. 
The second term of Eq.~(\ref{eq:He6_wf}) 
describes a dineutron which can be expanded largely from the $\alpha$ core changing its size. 
Here we also choose the parameters 
$\beta = 2, 3, \ldots, 8$ fm and five $b_n$ values for each $\beta$
as in the $^8$He DC wave functions.

\subsubsection{Quantitation of dineutron component}
\label{secII-A-5}
In the present work, we are mainly interested in the dineutron components in $^8$He. 
In order to see the one-dineutron component, 
we calculate the overlap of the $^8$He wave function (Eq.~(\ref{eq:He8_wf})) 
with the DC wave function of the $^6$He core and one dineutron
whose size ($b_n$) and spatial expansion ($\beta$) from the core change.  
The overlap of $^8$He is defined as
\begin{align}
& \mathcal{N}_{\rm DC}^{^8{\rm He}} (\beta, b_n) \nonumber \\
&\ =
|\langle \mathcal{P}^{0+}_{00} \Phi_{\rm DC} 
(\lambda = \lambda_0, d_{\lambda} = d_0, b_{\lambda} = b_{\alpha}; \beta, b_n) 
|\Psi_{^8 {\rm He}(0^+)} \rangle|^2.
\label{eq:overlap_DC_He8}
\end{align}
By calculating them with the various values of the parameters $b_n$ and $\beta$, 
we investigate what one-dineutron component 
the $0^+$ states of $^8$He contain. 
In the overlapped DC wave function, 
the $^6$He$(0^+_1)$ core is assumed to be composed of an $\alpha$ cluster and 
a $2n^*$ cluster with one configuration, 
which means that 
the parameters characterizing the dissociation and distance from the $\alpha$ core 
of the $2n^*$ cluster, 
$\lambda$ and $d_{\lambda}$, are fixed to a certain value $\lambda_0$ and $d_0$. 
These values are determined as 
the $\alpha+2n^*$ cluster wave function 
having $\lambda_0$ and $d_0$ gives the largest overlap
with $^6$He$(0^+_1)$ obtained by the present calculation. 
Then, the overlap of Eq.~(\ref{eq:overlap_DC_He8}) 
corresponds to considering the one-dineutron behavior 
around a kind of approximated $^6$He core. 
We also calculate the overlap of $^6$He$(0^+)$ obtained in the present calculation
with the DC wave function of $\alpha$ and one dineutron. 
\begin{equation}
\mathcal{N}_{\rm DC}^{^6{\rm He}} (\beta, b_n)  =
|\langle \mathcal{P}^{0+}_{00} \Phi_{\rm DC} (\beta, b_n) 
|\Psi_{^6 {\rm He}(0^+)} \rangle|^2. 
\label{eq:overlap_DC_He6}
\end{equation}
The overlap of $^6$He also contains two parameters of $b_n$ and $\beta$, 
which characterized the dineutron size 
and the expansion of the dineutron center of mass distribution from the $\alpha$ core. 
In Sec.~\ref{secIII}, 
we investigate the one-dineutron component in $^8$He$(0^+_1)$ 
and that in $^6$He$(0^+_1)$
and consider the analogy and difference of the one-dineutron behavior in them.

In order to estimate the component of two-dineutron condensation, 
we calculate the overlap with the two-dineutron condensate (2DC) wave function, 
$\mathcal{N}_{2 {\rm DC}}$, defined as follows.
\begin{equation}
\mathcal{N}_{\rm 2DC} (\beta, b_n) =
|\langle \Phi_{\rm 2DC} (\beta, b_n) | \Psi_{^8 {\rm He}(0^+)} \rangle|^2, 
\label{eq:overlap_2DC}
\end{equation}
where the 2DC wave function, $\Phi_{\rm 2DC} (\beta, b_n)$, 
describes the two-dineutron condensate system, 
which means two dineutrons occupying the same $S$-wave around an $\alpha$ with the same size. 
We perform the approximation for the description of two dineutrons in the $S$-wave, 
and explain the details of the 2DC wave function in Appendix~\ref{appendix1}. 
Analyzing the overlap with the 2DC wave function, $\mathcal{N}_{2 {\rm DC}} (\beta, b_n)$, 
we investigate the component that two dineutrons occupy the same $S$-wave 
in $^8$He$(0^+)$.

\subsection{Hamiltonian}
\label{secII-B}
In the present work, we use the Hamiltonian, 
\begin{equation}
H = T - T_G + V_{\rm cent} + V_{\rm LS} + V_{\rm Coul}, 
\label{eq:hamiltonian_DC}
\end{equation}
where $T$ and $T_G$ are the total kinetic energy and that of the center of mass motion. 
In the present framework, since all of the Gaussian size parameters are not the same 
in the DC wave function,
the center of mass motion cannot be removed exactly. 
Accordingly, we explicitly extract the expectation value of the center of mass kinetic energy 
from the total Hamiltonian as an approximation in Eq.~(\ref{eq:hamiltonian_DC}). 
$V_{\rm cent}$ and $V_{\rm LS}$ are the central and spin-orbit force,  
and we use the Volkov No.2 force as $V_{\rm cent}$ \cite{volkov65}
and the spin-orbit part of the G3RS potential as $V_{\rm LS}$ \cite{tamagaki68}. 
We fix the strength of the spin-orbit force to $v_{\rm LS} = 2000$ MeV in the present calculation
as in Ref.~\cite{itagaki00}.
$V_{\rm Coul}$ is the Coulomb force 
and we approximate it by the summation of seven Gaussians. 
By using the effective two-body interaction, 
it is difficult to perfectly reproduce the binding energies of all the subsystems simultaneously. 
Therefore, we use two kinds of parameter sets 
with different Majorana $(m)$, Bartlett $(b)$ 
and Heisenberg $(h)$ parameters in the central force. 
One set has $m=0.55, \ b=h=0.125$
and the other has $m=0.59, \ b=h=0.0$, 
which are labeled as ``m55'' and ``m59'' in the following, respectively. 
These sets are adjusted to reproduce the ground state energy of $^8$He
with respect to the threshold of $\alpha+4n$ as shown later. 
In the case of the m55 set, the interaction between two neutrons in the spin-singlet channel
is adequate to describe the unbound property of a two-neutron system, 
but the attraction between an $\alpha$ and a valence neutron is stronger 
than that in the case of 
$m = 0.60$ that reproduces the phase shift of the $\alpha$-$n$ scattering \cite{okabe79}. 
On the other hand, in the m59 set with $b=h=0.0$, 
the attraction between two neutrons in the spin-singlet channel is too strong
and two neutrons are bound contrary to experiments, 
while the $\alpha$-$n$ interaction is reasonable.
We compare the results using these sets 
to see how the dineutron correlation in $^8$He and in $^6$He depends on
the $\alpha$-$n$ or $n$-$n$ interaction. 
We comment that, 
when we use the parameter set of $m = 0.60$ and $b=h=0.125$ which reproduces well
the $\alpha$-$n$ scattering phase shift and the $n$-$n$ unbound feature, 
we get the underbound $^8$He system the same as the other works of $^8$He
\cite{enyo07, itagaki08}.

\subsection{Result}
\label{secII-C}

In this section, we discuss the structures of the ground and excited $0^+$ states of $^8$He
calculated with two sets of the interaction parameters. 
First, we show their fundamental characteristics such as the energies and the radii, 
and next we investigate the dineutron components in them. 
We also compare the structure of $^8$He$(0^+_1)$ with that of $^6$He$(0^+_1)$
to discuss the difference of the dineutron behavior between them.

\subsubsection{Energy spectra and characteristics}
\label{secII-C-1}
\begin{table*}[!t]
\caption{The structure properties of the $0^+_1$ state of $^6$He
the $0^+_{1,2}$ states of $^8$He; 
the rms radii of the matter and proton and neutron parts ($r_m, r_p$ and $r_n$), 
the expectation value of the neutron squared spin ($\langle S_n^2 \rangle$) 
and the neutron $0p_{3/2}$-component ($0p_{3/2}$), 
calculated with the m59 and m55 sets. 
The experimental values of rms radii of $^{6,8}$He$(0^+_1)$ are listed. 
((c) from Refs.~\cite{tanihata88, tanihata92} and (d) from Ref.~\cite{wang04}.)
For $^6$He$(0^+_1)$, the theoretical values of other calculations
with three-body models are also shown. 
((a) from Ref.~\cite{arai99} and (b) from Ref.~\cite{hagino05}.)}
\label{tab:characteristics}
\begin{ruledtabular}
\begin{tabular}{lcccccccc}
&&& $r_m$ (fm) & $r_p$ (fm) & $r_n$ (fm) & $\langle S_n^2 \rangle$ & $0p_{3/2}$ (\%) & \\ 
\hline
$^6$He$(0^+_1)$ 
& m55 && 2.30 & 1.91 & 2.47 & 0.12 & 67 & \\
& m59 && 2.35 & 1.96 & 2.53 & 0.08 & 59 & \\
& Refs.~\cite{arai99, hagino05} && 2.33$^{({\rm a})}$ & 1.76$^{({\rm a})}$ & 2.57$^{({\rm a})}$ & 
0.24$^{({\rm a})}$, 0.26$^{({\rm b})}$ & 83$^{({\rm b})}$ & \\
& EXP. && 2.33$-$2.48$^{({\rm c})}$ & 1.912 $\pm$ 0.018$^{({\rm d})}$ & 
2.59$-$2.61$^{({\rm c})}$ & & & \\
\hline
$^8$He$(0^+_1)$ 
& m55 && 2.37 & 1.87 & 2.51 & 0.63 & 73 & \\
& m59 && 2.49 & 1.93 & 2.65 & 0.41 & 55 & \\
& EXP. && 2.49$-$2.52$^{({\rm c})}$ & 1.76$-$2.15$^{({\rm c})}$ & 
2.64$-$2.69$^{({\rm c})}$ & & & \\
\hline
$^8$He$(0^+_2)$ 
& m55 && 4.85 & 2.26 & 5.44 & 0.14 & 4 & \\
& m59 && 4.67 & 2.21 & 5.24 & 0.09 & 3 & \\
\end{tabular}
\end{ruledtabular}
\end{table*}

\begin{figure}[!t]
\includegraphics[scale=0.65]{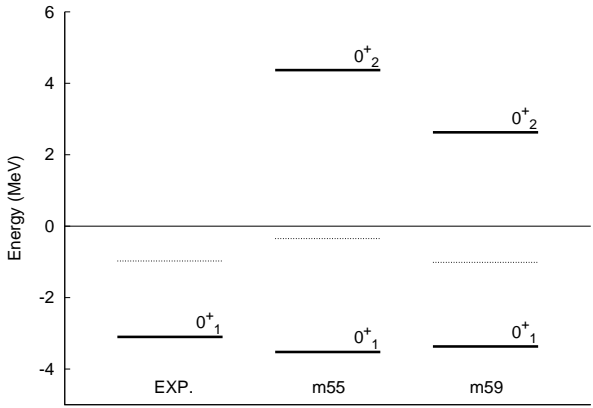}
\caption{Energy spectra of the $0^+$ states in experiments and the present calculation
measured from the threshold of $\alpha+4n$.
The threshold of $^6$He$+2n$ is shown with the dotted line in each spectrum.}
\label{fig:ene_spect}
\end{figure}

We calculate the energy spectra of the $0^+$ states and show them in Fig.~\ref{fig:ene_spect}
as well as the experimental one. 
The absolute values of the ground state of $^8$He 
(labeled as $0^+_1$ in Fig.~\ref{fig:ene_spect}) are 
$-31.40$ MeV (EXP.), $-31.13$ MeV (m55) and $-30.98$ MeV (m59), 
and those of the $\alpha+4n$ threshold are 
$-28.30$ MeV (EXP.) and $-27.61$ MeV (m55 and m59). 
In addition to the $0^+_1$ state, 
we obtain the $0^+_2$ state which is a resonance state
above the $\alpha+4n$ and $^6$He$+2n$ threshold energies, 
and strictly speaking, it couples with the continuum states. 
In the analysis of the dineutron component shown in Sec.~\ref{secII-C-3}
and that with a pseudo potential method, 
we identify this states as the resonance. 
The detail of the pseudo potential method is explained in Appendix~\ref{appendix2}. 

We also calculate the energy of the ground state of $^6$He,
and plot its relative energy to the $\alpha+2n$ threshold
(it is equivalent to the relative energy 
between the $^6$He$+2n$ threshold and the $\alpha+4n$ threshold) 
with the dotted lines in Fig.~\ref{fig:ene_spect}. 
Their absolute values are 
$-29.27$ MeV (EXP.), $-27.95$ MeV (m55) and $-28.62$ MeV (m59). 
As for the energy of the ground state of $^6$He,
the present calculation reproduces well the experimental value in the m59 case, 
but it is slightly underbound in the m55 case. 

\begin{figure}[!t]
\includegraphics[scale=0.65]{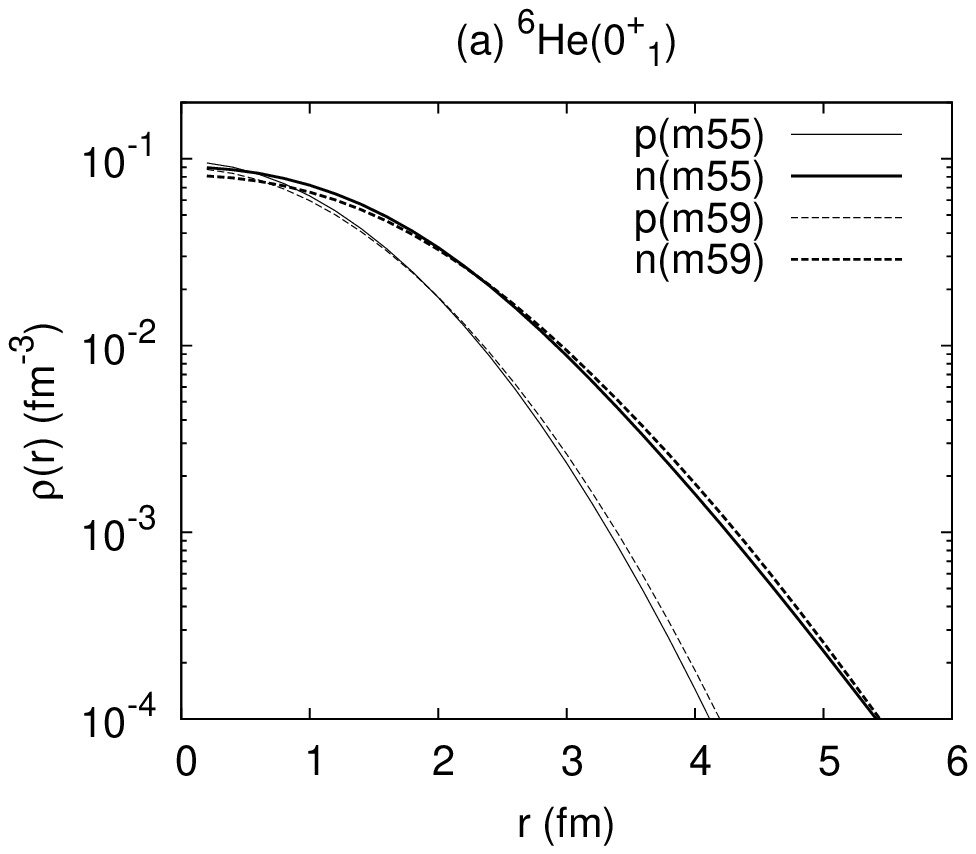} \\
\includegraphics[scale=0.65]{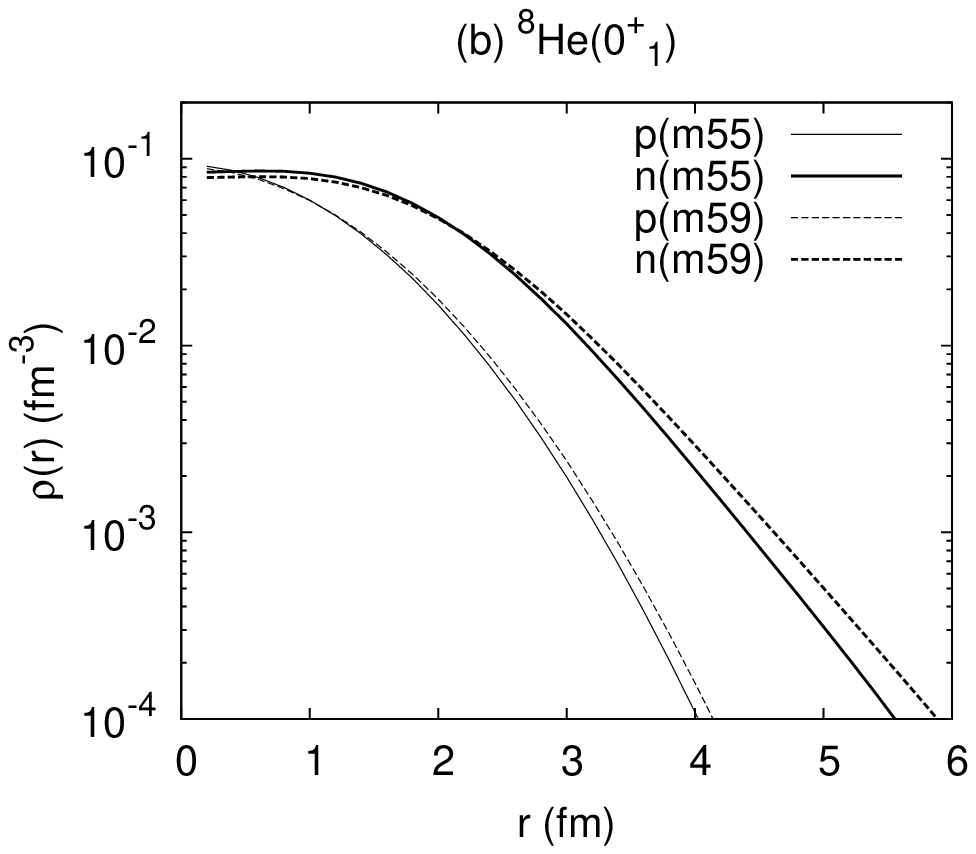} 
\caption{The proton and neutron one-body densities of $^{6,8}$He$(0^+_1)$. 
The solid and dashed lines correspond to the densities 
calculated with the m55 and m59 interaction sets, respectively. 
The thick and thin ones are the neutron and proton densities.}
\label{fig:density_1st}
\end{figure}
In order to see the degree of spatial expansion of these states, 
we calculate the root-mean-square (rms) matter, proton and neutron radii 
($r_m, r_p$ and $r_n$) of $^6$He$(0^+_1)$ and $^8$He$(0^+_{1,2})$ 
shown in Table.~\ref{tab:characteristics}. 
It should be noticed that, 
in the calculation of the $^8$He$(0^+_2)$ radii, 
we superpose only the truncated set of basis wave functions of Eq.~(\ref{eq:He8_wf}), 
as $d_{\lambda} \leq 6$ fm for $\Phi_{\rm DC}$ with $\lambda = 0.0$
and $\beta \leq 6$ for both $\Phi_{\rm DC}$ with $\lambda = 0.0, 0.4$. 
It is because the resonance state of $^8$He$(0^+_2)$ is likely to couple to continuum states
so that the radii diverge if $d_{\lambda}$ and $\beta$ are large 
as chosen in the full bases shown in Table~\ref{tab:parameter_He8}. 
Therefore, we evaluate the $^8$He$(0^+_2)$ radii with the truncated bases. 
Other properties of $^8$He$(0^+_2)$ 
such as the binding energy and the dineutron component discussed below 
are calculated with the full set of the bases in Table~\ref{tab:parameter_He8}. 
We have checked that the results except for the radii do not change so much 
with and without the truncation. 

The calculated values of the $^{6,8}$He$(0^+_1)$ radii agree with the experimental values, 
i.e. the reproduction of the much larger neutron radii than the proton one is very good. 
We also calculate the proton and neutron one-body densities, $\rho_{p,n}(r)$, 
in $^{6,8}$He$(0^+_1)$. 
As shown in Fig.~\ref{fig:density_1st}, 
the valence neutrons are distributed far from the $\alpha$ core
to form the neutron-halo structure in $^6$He$(0^+_1)$ and 
the neutron-skin structure in $^8$He$(0^+_1)$. 
There exists a slight difference between the result of the m55 set and that of the m59 set
in the neutron density in $^8$He. 
The neutron expansion is larger in the m59 set, 
and it is partly because of the weaker $\alpha$-$n$ attraction, 
and partly because of the more enhanced dineutron correlation as shown later,
due to the strong $n$-$n$ interaction. 
The characteristics of the one-body densities in $^6$He$(0^+_1)$ and $^8$He$(0^+_1)$
are qualitatively consistent with the other calculations,  
for instance, that using an extended three-body model in Ref.~\cite{arai99}. 
Quantitatively, however, the neutron-halo tail of $^6$He$(0^+_1)$ 
in the region of $r \gtrsim 5$ fm
is smaller in the present result than that in Ref.~\cite{arai99}. 
It may be because the component that only one neutron is expanded largely from the core
is not sufficiently taken into account in the present model space. 

$^8$He$(0^+_2)$ state has 
the much larger radii than $^8$He$(0^+_1)$ as shown in Table~\ref{tab:characteristics}.
Compared to $^8$He$(0^+_1)$, the neutron radius is remarkably large, 
though the proton radius also increases slightly due to the recoil effect 
between the $\alpha$ and the valence neutrons.
Such a huge neutron radius demonstrate the gas-like feature 
of two dineutrons expanded largely around an $\alpha$. 
As discussed in the previous works \cite{yamada08, yamada08_2}, 
such an excited state with developed clusters tends to have 
a significant monopole transition strength to the ground state. 
We calculate the neutron monopole transition strength of $^8$He$(0^+_2)$ to $^8$He$(0^+_1)$, 
and get the values of the strength, 9.0 fm$^2$ and 10.3 fm$^2$ in the m55 and m59 sets, respectively. 
They exhaust the first-order energy weighted sum rule by 22\% and 21 \%, 
indicating that the neutron monopole transition strength between the $0^+_1$ and $0^+_2$ states 
is certainly large as expected
and that a monopole transition would be a useful probe
to experimentally observe the $0^+_2$ state 
containing two dineutrons expanded largely.

In order to discuss how much the spin-singlet two-neutron pairs
break into the $j$-$j$ coupling state with the $0p_{3/2}$-shell configurations, 
we show the expectation value of the squared spin of the neutron part, 
$\langle S_n^2 \rangle$, in Table~\ref{tab:characteristics}. 
We also show in Table~\ref{tab:characteristics} 
the component of the ideal $0^+$ states of the H.O. $0p_{3/2}$-shell configurations, 
$(0p_{3/2})^2_{J^{\pi}=0^+}$ in $^6$He$(0^+_1)$ 
and $(0p_{3/2})^4$ in $^8$He$(0^+_{1,2})$, 
that is the overlap, $|\langle (0p_{3/2})^2|\Psi_{^6{\rm He}(0^+_1)} \rangle|^2$
and $|\langle (0p_{3/2})^4|\Psi_{^8{\rm He}(0^+_{1,2})} \rangle|^2$, respectively. 
In the present framework, 
we assume an $\alpha$ core which gives no contribution to $\langle S_n^2 \rangle$, 
so a non-zero $\langle S_n^2 \rangle$ value indicates the breaking of spin-singlet pair(s)
of valence neutrons around the $\alpha$ core. 
In the $(0p_{3/2})^2$ limit of $^6$He, $\langle S_n^2 \rangle = 2/3$, 
and in the $(0p_{3/2})^4$ limit of $^8$He, $\langle S_n^2 \rangle = 4/3$. 
In $^6$He$(0^+_1)$, 
the spin expectation value, $\langle S_n^2 \rangle$, is very small 
compared with the value of the $(0p_{3/2})^2$ limit. 
It means that the component of a spin-singlet pair of two valence neutrons, 
the so-called dineutron component, is dominant in the ground state of $^6$He. 
A slight non-zero value of $\langle S_n^2 \rangle$ means that 
the dissociation of the dineutron has a little effect to $^6$He$(0^+_1)$. 
A spin-singlet pair of two neutrons in the $0p$-shell around an $\alpha$ core
has the overlap of $2/3$ with the $|(0p_{3/2})^2 \rangle$ state, 
and our calculated values in Table~\ref{tab:characteristics} are almost consistent with this value. 
It means that $^6$He$(0^+_1)$ contains, as the dominant one, 
the zero-spin component of two valence neutrons
in both cases of the m55 and m59 sets. 

$^8$He$(0^+_1)$ has the significant $\langle S_n^2 \rangle$ value 
indicating the breaking of the spin-singlet two-neutron pair(s). 
Such breaking of the dineutrons
originates in the stronger binding in $^8$He than $^6$He. 
The stronger binding induces the more dissociation of spin-singlet pairs at the nuclear surface
so as to gain the spin-orbit energy. 
As a result, $^8$He$(0^+_1)$, especially in m55,
has a larger spin expectation value compared with $^6$He$(0^+_1)$, 
which means that the shell-model configuration is more important 
in $^8$He$(0^+_1)$ than in $^6$He$(0^+_1)$.
However, these $\langle S_n^2 \rangle$ values are 
less than half as much as that of the $(0p_{3/2})^4$ limit, 
so the dineutron component is also significant as well as the shell-model one. 
In other words, the $j$-$j$ coupling feature and the dineutron correlation
compete in the ground state of $^8$He.   

On the other hand, 
$^8$He$(0^+_2)$ has much smaller $\langle S_n^2 \rangle$ value
than that of the $0^+_1$ state
and contains only a few percent $(0p_{3/2})^4$ component. 
It is because two dineutrons develop in the $0^+_2$ state.
As is shown later, 
the $0^+_2$ state shows the feature of two dineutron condensation, 
which is characterized by the structure of weakly interacting spin-zero clusters 
with little correlation each other. 

Such values as the spin expectation value or the shell-model component discussed here
depend on the strength of the interaction somehow. 
However, there are no qualitative differences 
between the results of m59 and m55 shown above, 
and these results are consistent with the ones in the previous works
\cite{enyo07, arai99, hagino05}. 
Then, it can be said that $^{6,8}$He is well-described in the present framework. 
In the following, we closely investigate the dineutron components in $^8$He$(0^+_{1,2})$, 
which is the main issue of this work.  
We investigate the one- and two-dineutron contribution in $^8$He 
and compare the dineutron behavior in $^8$He$(0^+_1)$ and that in $^6$He$(0^+_1)$. 
In addition, we discuss the dineutron condensation in $^8$He$(0^+_2)$. 

\begin{figure*}[!t]
\begin{center}
\begin{tabular}{cc}
\includegraphics[scale=0.65]{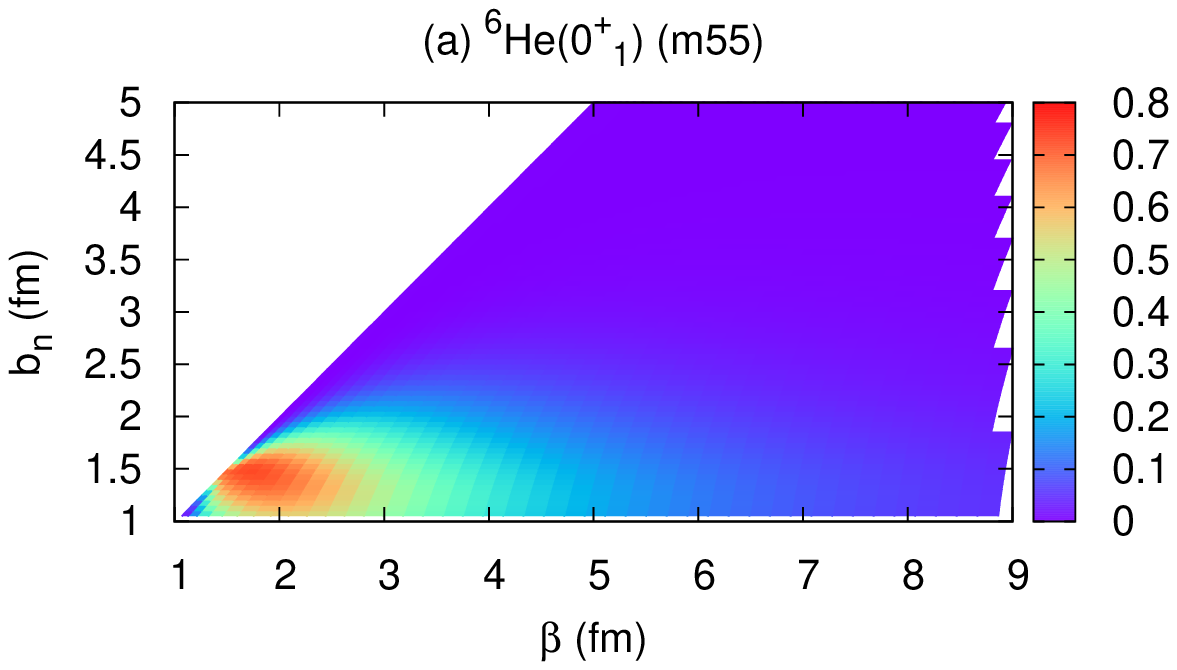} &
\includegraphics[scale=0.65]{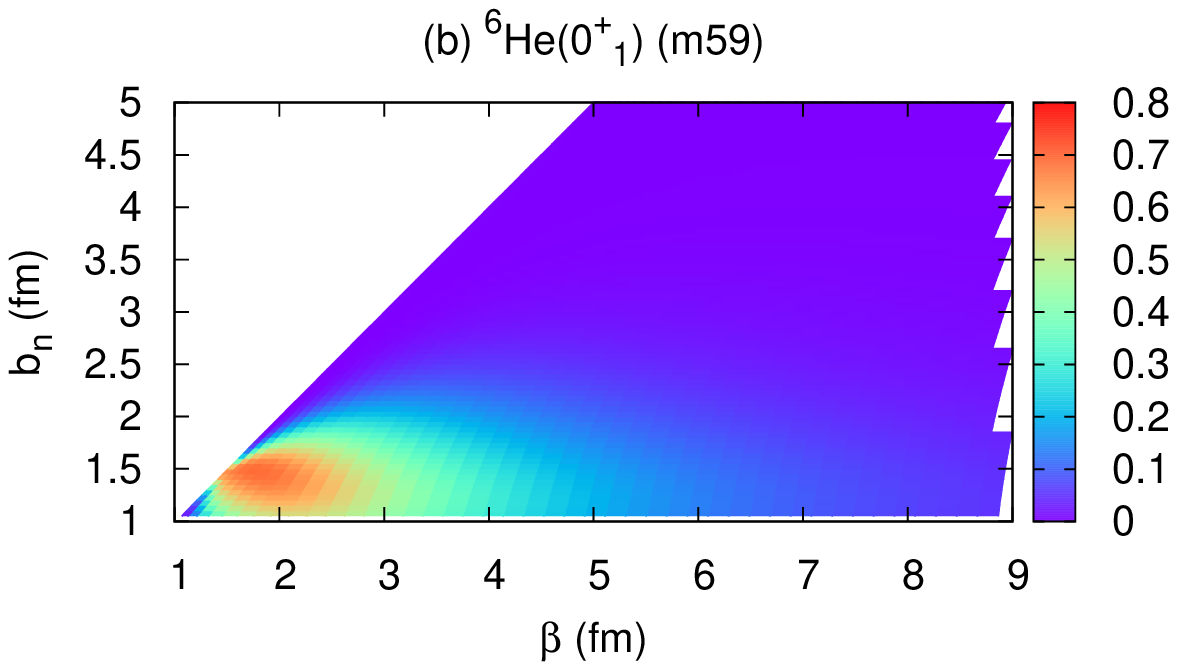} \\
\includegraphics[scale=0.65]{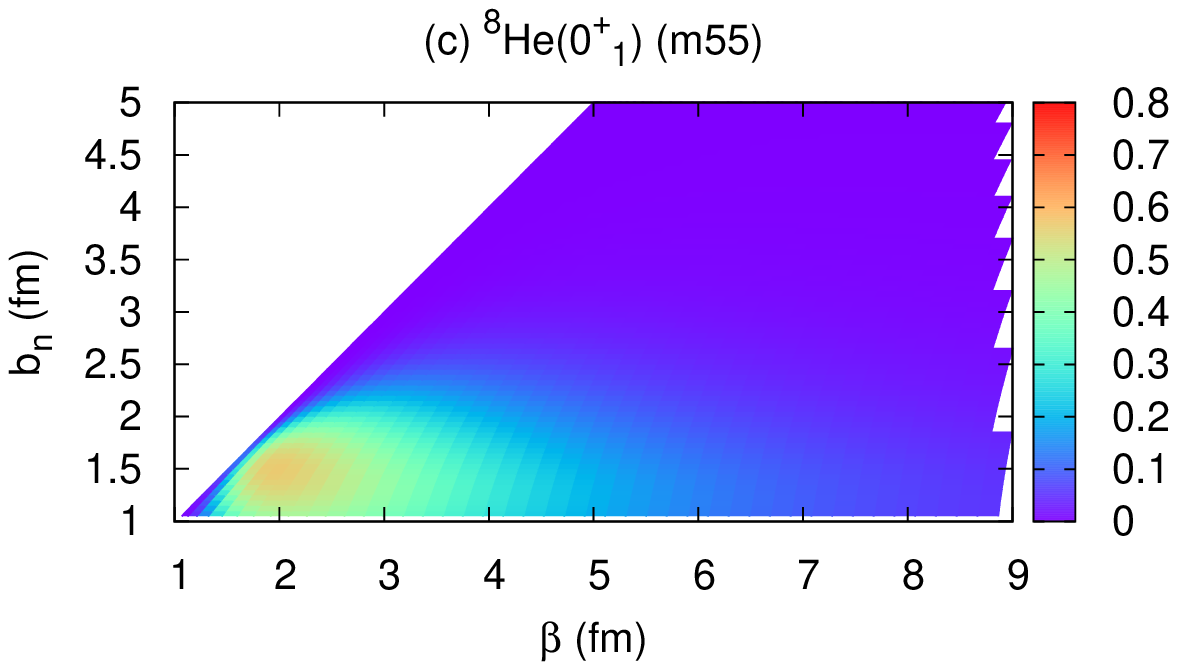} &
\includegraphics[scale=0.65]{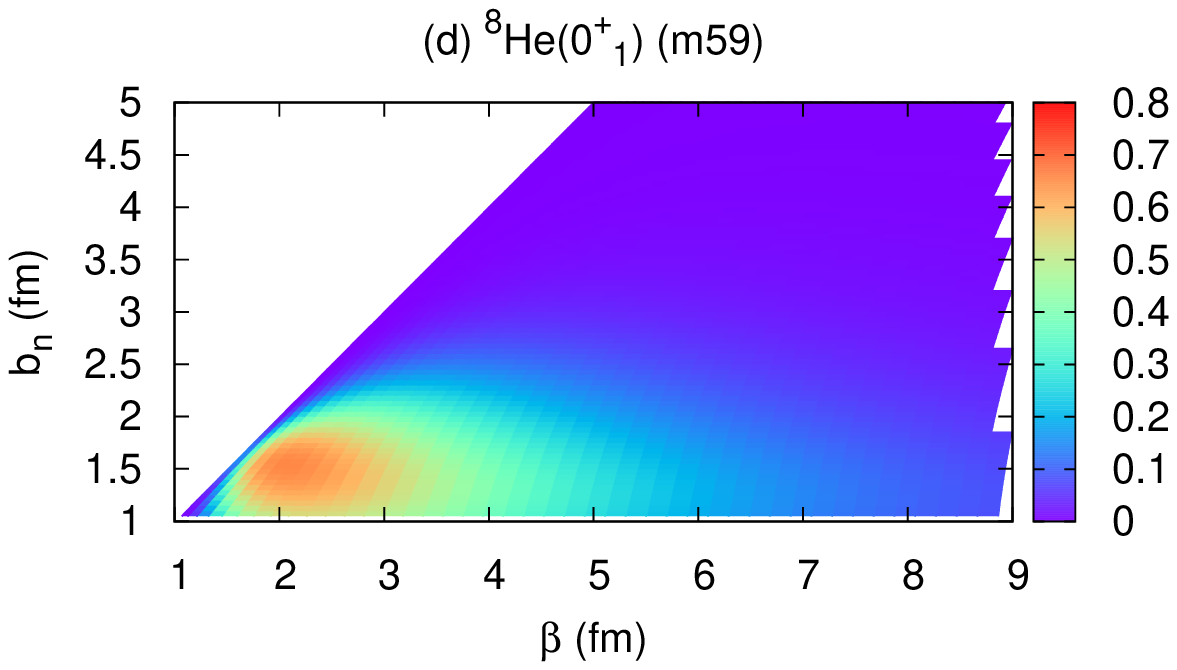} \\
\end{tabular}
\caption{(Color online) The overlaps of the $^{6,8}$He$(0^+_1)$ states
with the DC wave function on the $\beta$-$b_n$ plane.
The upper figures are of $^6$He, and the lower ones are of $^8$He. 
The left ones are calculated with the m55 set, and the right ones are with the m59 set.}
\label{fig:overlap_B_b_DC_1st}
\end{center}
\end{figure*}

\subsubsection{Component of one-dineutron in the $S$-wave 
in $^8$He$(0^+_1)$ and $^6$He$(0^+_1)$}
\label{secII-C-2}
We consider the contribution of the dineutron correlation 
to the ground states of $^8$He and $^6$He. 
As explained in Sec.~\ref{secII-A-5}, 
we calculate the overlaps, $\mathcal{N}_{\rm DC}^{^{6,8}{\rm He}}$ 
(Eq.~(\ref{eq:overlap_DC_He8}) and (\ref{eq:overlap_DC_He6})), 
of the obtained ground state wave functions 
with a DC wave function assuming that the system consists of a core and one dineutron
to see the component of one-dineutron in the $S$-orbit. 
As explained in Sec.~\ref{secII-A-5}, 
in the calculation of $\mathcal{N}_{\rm DC}^{^8{\rm He}}$, 
we approximate the $^6$He core in the overlapped $^8$He DC wave function 
by a single $\alpha+2n^*$ cluster wave function,
$\Phi_{\alpha+2n^*}$, having a fixed parameters $\lambda_0$ and $d_0$ 
which are determined so as to give the largest overlap 
with the $^6$He ground state wave function, $\Psi_{^6{\rm He}(0^+_1)}$. 
The values are $(\lambda_0, d_0) = (0.3, 3.0)$ in the m59 set, 
and $(0.4, 2.5)$ in the m55 set. 
The $\Phi_{\alpha+2n^*}(\lambda_0,d_0)$ wave function with these values
gives the largest overlap of $\sim 90$\% 
with the present $^6$He$(0^+_1)$ wave function in each case, 
so it is a good approximation of $^6$He$(0^+_1)$.
We calculate the overlap, $\mathcal{N}_{\rm DC}^{^8{\rm He}}$, 
with the DC wave function of one dineutron around the approximated $^6$He core
as the function of the dineutron size, $b_n$, and 
the spatial expansion, $\beta$, of the dineutron center of mass distribution around the core, 
and we discuss what component of a dineutron is contained in the $0^+_1$ states. 
The overlaps of $^{6,8}$He$(0^+_1)$ with the DC wave function
are plotted on the $\beta$-$b_n$ plane
in Figs.~\ref{fig:overlap_B_b_DC_1st}. 

First, we consider how the one-dineutron component depends on the interaction parameter 
comparing the results of the m55 and m59 sets.
In the $^6$He case (Figs.~\ref{fig:overlap_B_b_DC_1st}(a) and (b)), 
the component of the compact dineutron at the nuclear surface is dominant, 
as seen in the peak at $(\beta, b_n) \sim (1.8, 1.5)$. 
Along the $b_n \sim 1.5$ fm line, 
a long tail of the significant amplitude is seen in the $\beta \gtrsim 4$ fm region,  
which indicates that the dineutron formed at the very nuclear surface
can be expanded outward from the core keeping its size compact. 
Although the peak amplitude is a little larger in m55 (the strong $\alpha$-$n$ interaction)
than in m59 (the strong $n$-$n$ interaction), 
such a tendency of the peak and tail structure 
along the $b_n \sim 1.5$ fm line is not different in both the results of $^6$He$(0^+_1)$. 
It indicates that the dineutron formed at the nuclear surface 
contributes to the neutron-halo region because of the extremely weak binding in $^6$He, 
independently of the interaction between an $\alpha$ and a valence neutron
or the one between valence neutrons.
Also in the ground state of $^8$He (Figs.~\ref{fig:overlap_B_b_DC_1st}(c) and (d)), 
the peak amplitude at $(\beta,b_n) \sim (2.0, 1.6)$
indicates again the component of the compact dineutron at the surface. 
Contrary to $^6$He, 
the dineutron correlation in $^8$He is somewhat sensitive to the interaction 
due to the competition between the $j$-$j$ coupling feature and the dineutron correlation. 
Comparing the results of $^8$He$(0^+_1)$ obtained 
with the m55 and m59 sets, Figs.~\ref{fig:overlap_B_b_DC_1st}(c) and (d),
it is seen that
the peak and tail amplitudes are reduced in m55 where the $\alpha$-$n$ attraction is strong. 
It means that, because  the dineutrons are attracted strongly to the core 
and dissociated in m55, 
the dineutron component is reduced in the whole region.
In addition, since the interaction between two neutrons coupled to spin-singlet 
is weaker in m55 than that in m59, 
the dineutron correlation tends to be suppressed in m55. 

Next, we discuss the difference of the one-dineutron contribution 
between $^8$He$(0^+_1)$ and $^6$He$(0^+_1)$. 
As seen in the results calculated with m55 (Figs.~\ref{fig:overlap_B_b_DC_1st}(a) and (c)), 
the component of the compact dineutron at the very nuclear surface ($(\beta, b_n) \sim (2.0, 1.5)$) 
is reduced in $^8$He compared with that in $^6$He. 
It is because the dineutron and shell-model components compete each other in $^8$He
while the dineutron component is dominant in $^6$He. 
The neutron $(0p_{3/2})^4$ sub-shell closed configuration in $^8$He is 
more stable and favorable energetically, 
so the binding of the total system is stronger in $^8$He than that in $^6$He. 
As a result, the dineutron is pulled nearer the core
and dissociated due to the spin-orbit force at the nuclear surface in $^8$He. 
On the other hand, 
the tail amplitude in the $\beta \gtrsim 4$ fm region 
where the dineutron is distributed far from the core
is not so different between $^8$He and $^6$He. 
The spin-orbit interaction works just at the nuclear surface
and it does not necessarily dissociate a spin-singlet pair of two neutrons 
at the region far from the core. 
Therefore, at the outer region from the surface, 
two of the valence neutrons form a compact dineutron to be expanded spatially far from the core. 
We conclude that the one-dineutron component far from the core can be important 
in the ground states of both $^6$He and $^8$He, 
though it is somewhat reduced at the surface of $^8$He.

\begin{figure}[h]
\includegraphics[scale=0.65]{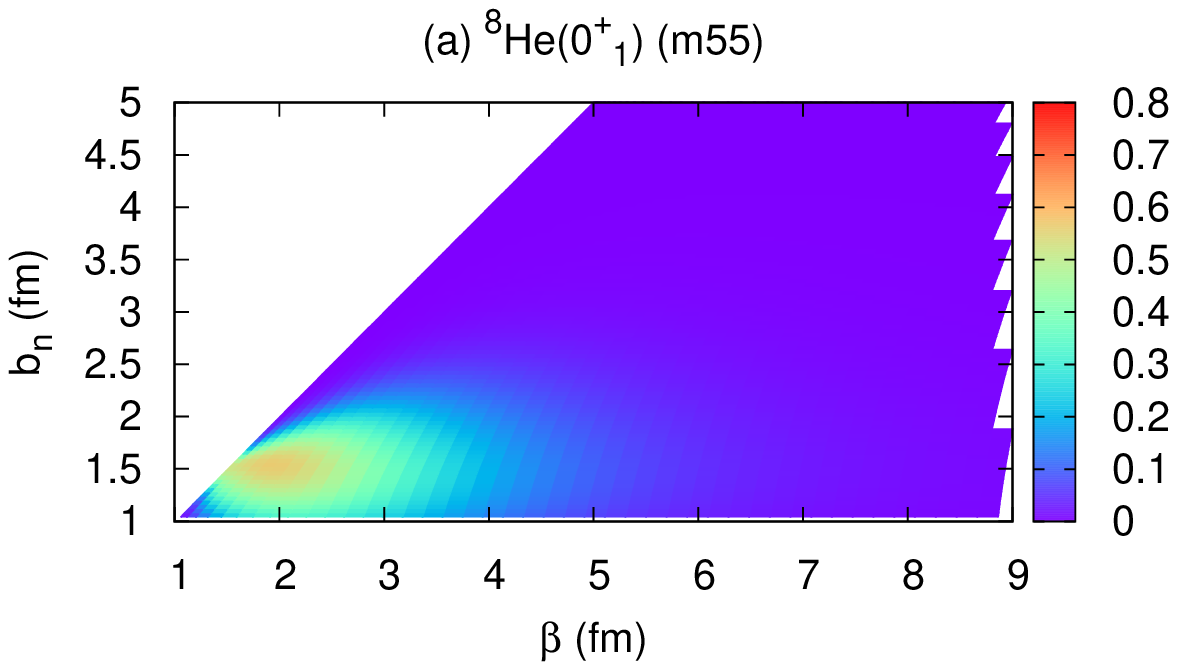} \\
\includegraphics[scale=0.65]{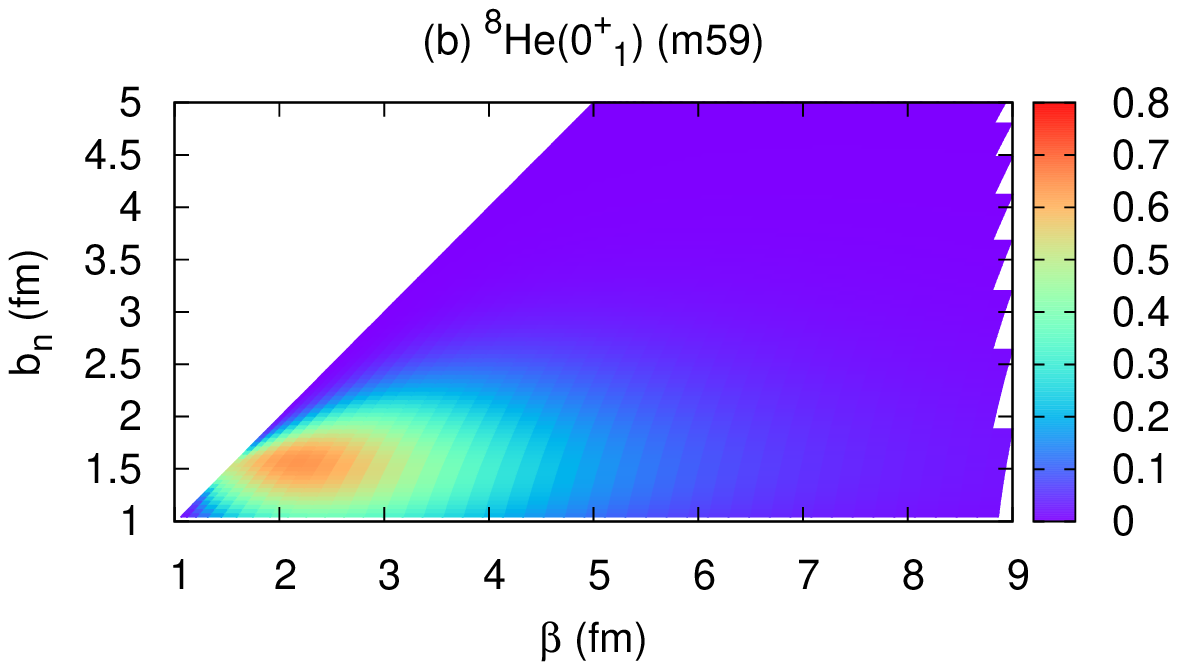} 
\caption{(Color online) The overlaps of the $^8$He$(0^+_1)$ state
with the 2DC wave function on the $\beta$-$b_n$ plane.
In the figures (a) and (b), we show the ones of the $0^+_1$ state 
calculated with the m55 and m59 parameter set, respectively.}
\label{fig:overlap_B_b_2DC_1st}
\end{figure}
In $^8$He, we are also interested in 
the component of two dineutrons in the same $S$-wave around the core, 
that is the two-dineutron condensation. 
In order to see the component of the two-dineutron condensation in $^8$He$(0^+)$, 
we calculate the overlap, $\mathcal{N}_{2{\rm DC}}$ (Eq.~(\ref{eq:overlap_2DC})), 
with the 2DC wave function explained in Sec.~\ref{secII-A-5} and Appendix~\ref{appendix1}.
We show in Figs.~\ref{fig:overlap_B_b_2DC_1st} 
the overlap, $\mathcal{N}_{2{\rm DC}}(\beta, b_n)$, of $^8$He$(0^+_1)$
as the function of the dineutron size, $b_n$, and 
the spatial expansion, $\beta$, of the dineutron center of mass distribution around the origin. 
In both the results of the m55 and m59 sets, 
it is found that two dineutrons are formed at the very nuclear surface, 
as shown by the peak at $(\beta, b_n) \sim (2.0, 1.6)$. 
This component of two dineutrons at the surface of the $\alpha$ core is observed 
also in the overlap, $\mathcal{N}_{\rm DC}^{^8{\rm He}}$, 
of the one-dineutron component having the maximum peak amplitude 
around the corresponding $(\beta, b_n)$
as shown in Figs.~\ref{fig:overlap_B_b_DC_1st}(c) and (d). 
However, we should emphasize that
the tail amplitude of two independent dineutrons in the $\beta \gtrsim 4$ fm region 
is largely suppressed, 
contrary to the long tail in the overlap, $\mathcal{N}_{\rm DC}^{^8{\rm He}}$, 
of the one-dineutron component in Figs.~\ref{fig:overlap_B_b_DC_1st}(c) and (d). 
It means that the component of two dineutrons distributed far from the core is minor
in the ground state of $^8$He. 
The suppression of the two-dineutron component in the large $\beta$ region is remarkable
especially in the m55 case where the $\alpha$-$n$ attraction is strong, 
while the amplitudes of the peak and the tail part along the $b_n \sim 1.5$ line 
are larger in the m59 case. 
It indicates that 
the dineutron components are more enhanced in the m59 case than in the m55 case 
due to the strong $n$-$n$ interaction and the weak $\alpha$-$n$ interaction
as discussed in the part of the one-dineutron component. 

As a conclusion, in the ground state of $^8$He, 
the $j$-$j$ coupling feature of the $(0p_{3/2})^4$ sub-shell closure 
and the dineutron correlation compete with each other. 
At the nuclear surface, 
the dineutron correlation still gives significant contribution
to form not only one but also two dineutrons. 
The distribution of one of two dineutrons can be expanded largely 
to cause the one-dineutron tail from the core,
but the spatial expansion of two dineutrons far from the $\alpha$ core 
does not occur in the $^8$He ground state.

\subsubsection{Dineutron condensation in $^8$He($0^+_2$)}
\label{secII-C-3}

\begin{figure}[h]
\includegraphics[scale=0.65]{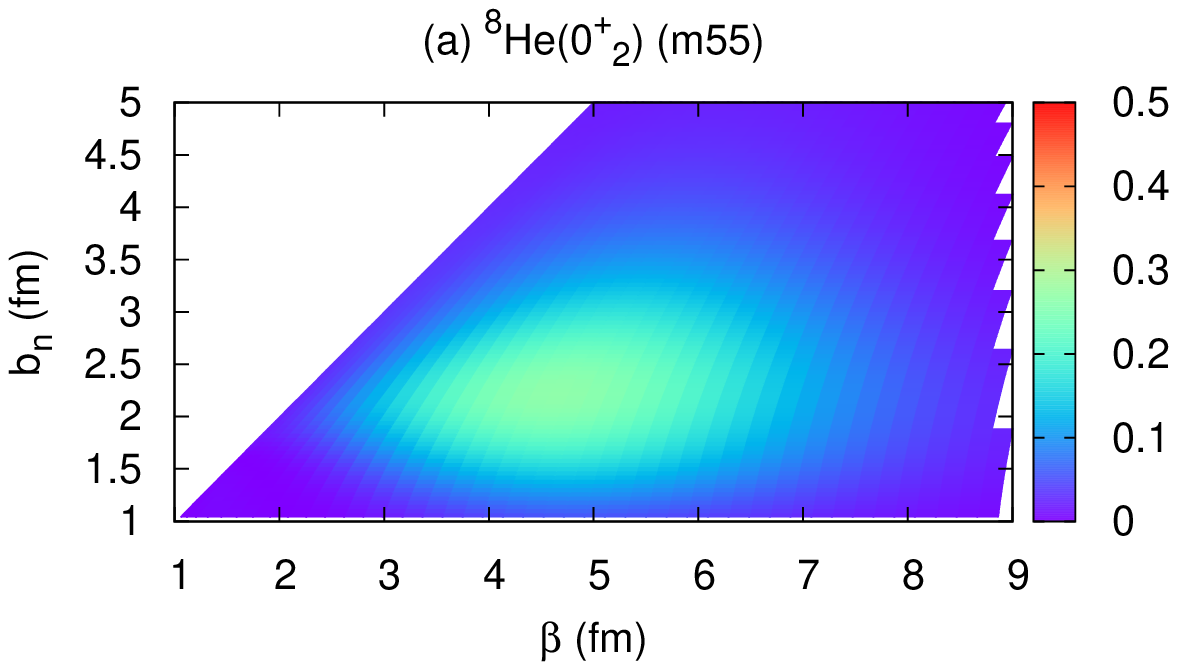} \\
\includegraphics[scale=0.65]{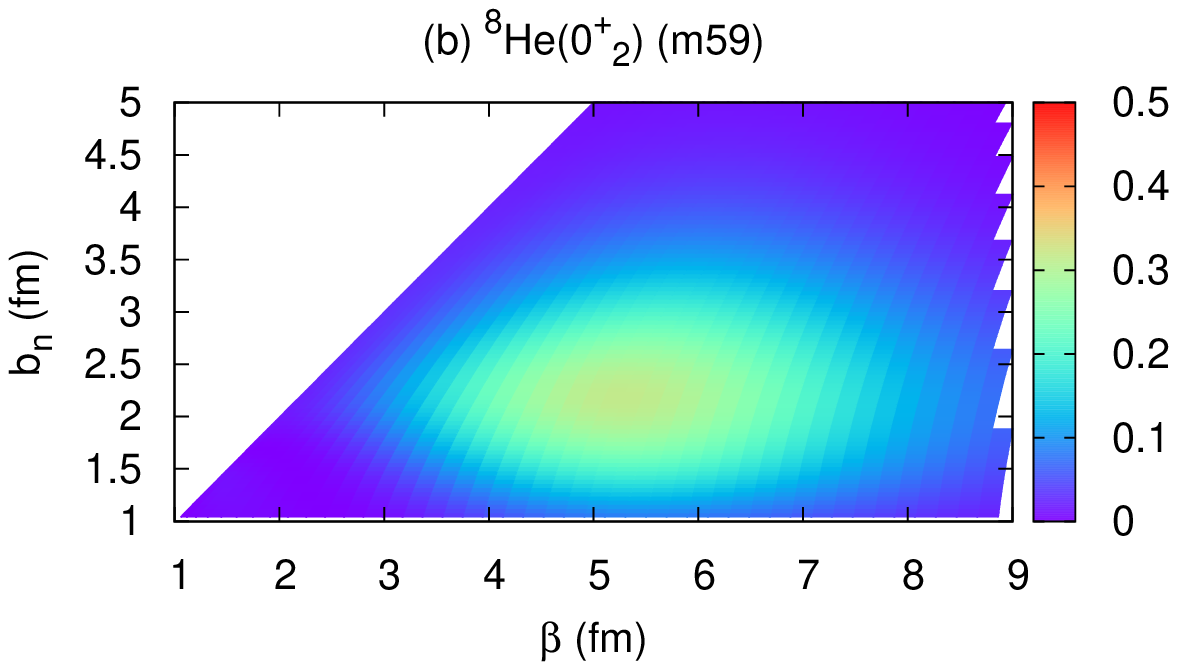} 
\caption{(Color online) The overlaps of the $^8$He$(0^+_2)$ state
with the 2DC wave function on the $\beta$-$b_n$ plane.
In the figures (a) and (b), we show the one of the $0^+_2$ state 
calculated with the m55 and m59 parameter set, respectively.}
\label{fig:overlap_B_b_2DC_2nd}
\end{figure}
In this section, we discuss the component of the two-dineutron condensation
in $^8$He$(0^+_2)$ which is located above the $\alpha+4n$ threshold by a few MeV
(Fig.~\ref{fig:ene_spect}). 
We calculate the overlap with the 2DC wave function, $\mathcal{N}_{2{\rm DC}}(\beta, b_n)$,  
to see the component of two dineutrons in the $0^+_2$ state
and plot the values on the $\beta$-$b_n$ plane in Figs.~\ref{fig:overlap_B_b_2DC_2nd}. 
The $0^+_2$ state in the m55 case  (Fig.~\ref{fig:overlap_B_b_2DC_2nd}(a))
has a significant amplitude of two dineutrons
in the wide region of $3 \lesssim \beta \lesssim 7$ fm and $1.5 \lesssim b_n \lesssim 3$ fm,
similarly in the m59 case (Fig.~\ref{fig:overlap_B_b_2DC_2nd}(b)). 
The broad width in the $\beta$ direction 
corresponds to two dineutrons expanded largely from the core 
and moving freely from the other clusters. 
In addition, 
the broad one in the $b_n$ direction means that the size of two neutrons can be fluctuated
from as compact size as an $\alpha$ ($b_n \sim 1.5$ fm)
to largely swelled one ($b_n \sim 3$ fm). 
Such broad peak may correspond to the dineutron condensation, 
which is characterized by a gas-like structure of loosely interacting clusters from each other. 
The dineutron correlation becomes somewhat weaker
because the nuclear density is dilute in such a state. 
In the ground state, valence neutrons are distributed at the nuclear surface, 
and two neutrons can correlate strongly to form a compact dineutron
because of the moderate surface density.  
On the other hand, in the excited state such as the $0^+_2$ state, 
valence neutrons are spatially expanded so largely from the $\alpha$ core that 
they feel a very low density. 
Thus the dineutron correlation becomes weaker 
and, as a result, its size becomes much larger 
than the typical dineutron size, $b_n \sim 1.5$ fm in the $^{6,8}$He$(0^+_1)$ states. 
Nevertheless, they form a subsystem of a spin-singlet two-neutron pair
considered as a kind of cluster, that is a dineutron cluster 
whose size is still smaller than the size of the total system. 
Such a change of the dineutron correlation depending on the density
should be related to the BCS-BEC crossover phenomenon 
of the dineutron correlation in nuclear matter \cite{matsuo06} and
in finite nuclei \cite{hagino07}. 
Comparing the component in the m59 (the strong $\alpha$-$n$ interaction) 
and m55 (the strong $n$-$n$ interaction) case, 
the amplitude is certainly larger 
and the peak region is shifted to the slightly larger $\beta$ region in m59 than in m55. 
These differences come from 
the fact that the dineutron correlation is enhanced in the m59 case
due to the stronger $n$-$n$ interaction. 
In addition, the ground state in the m59 set contains more large two-dineutron component 
in the further region than in the m55 
(comparing the tail components of $\beta \gtrsim 4$ fm in Fig.~\ref{fig:overlap_B_b_2DC_1st}(a) and (b))
so that the orthogonality to each ground state also might have some effect
to the dineutron correlation in the excited state. 

In the previous work on the dineutron correlation in $^{10}$Be system, 
we suggested the similar condensate state of $\alpha+\alpha+$dineutron 
in an excited state at a few MeV higher than the $\alpha+\alpha+2$n threshold \cite{kobayashi12}. 
The dineutron behavior in $^8$He$(0^+_2)$ is similar to 
that in the $\alpha+\alpha+$ dineutron condensate state in $^{10}$Be. 
It is expected that not only the $\alpha$ condensate states
which have been well-discussed so far, 
but also the condensate states containing some $\alpha$ and dineutron clusters
would exist above the corresponding threshold.

\section{Formation mechanism of dineutron condensation}
\label{secIII}

In this section, we discuss about the formation mechanism of the dineutron condensate state
 of $^8$He$(0^+_2)$ suggested in the previous section 
by considering an ideal situation of two dineutrons moving in the $S$-wave around the $\alpha$ core. 
For this aim, we introduce a new framework, 
the $\alpha$ and dineutron condensate wave function,
which is specialized to describe the dineutron condensation in finite nuclei. 
We discuss the behavior of two dineutrons around an $\alpha$ core, 
and compare it with the case of a triton core 
to show that the potential from the core is essential for the dineutron condensation.

\subsection{Framework}
\label{secIII-A}
In order to describe the $\alpha$ condensation in $4N$ nuclei 
such as $^{12}$C$(0^+_2)$, the so-called Hoyle state, 
the $\alpha$ condensate wave function
called as the THSR wave function, has been proposed by Tohsaki {\it et al}..
It has been shown in Refs.~\cite{tohsaki01, funaki03, funaki05, funaki09}
that the Hoyle state can be described well 
by a single $\alpha$ condensate wave function. 
The $\alpha$ condensate wave function has so simple a form with only a few parameters
that it is useful to clarify the mechanism of the $\alpha$ condensation.  
Following the $\alpha$ condensate wave function, 
we formulate a new framework for a condensate state of dineutrons with $\alpha$(s), 
which we call the $\alpha$ and dineutron condensate wave function,
in short the $\alpha$DC wave function hereafter. 
We apply it to discuss the mechanism of the dineutron condensation in the $\alpha+2n+2n$ system. 
Here we explain the $\alpha$DC wave function for the $\alpha+2n+2n$ system 
used in this section. 
The $\alpha+2n+2n$ $\alpha$DC wave function describes a system 
containing one $\alpha$ and two $2n$ clusters as
\begin{align}
& \Psi_{\alpha{\rm DC}}(B_{\alpha},b_{\alpha}, B_{2n},b_{2n}) \nonumber \\
&\ = \frac{1}{\sqrt{8!}} \ 
\mathcal{A} \left\{ \ \int d^3 \boldsymbol{R}_1
\exp \Big[ - \frac{\boldsymbol{R}_1^2}{B_{\alpha}^2} \Big]  
\Phi_{\alpha} (\boldsymbol{R}_1, b_{\alpha}) \right.  \nonumber \\ 
&\ \hspace{1em}
\times \left. 
\prod_{c = 2}^{3} \left( \int d^3 \boldsymbol{R}_c 
\exp \Big[ - \frac{\boldsymbol{R}_c^2}{B_{2n}^2} \Big]  
\Phi_{2n} (\boldsymbol{R}_c, b_{2n}) \right) \right \}. \label{eq:aDC_wf}
\end{align}
$\Phi_{\alpha}$ and $\Phi_{2n}$ are the ordinary $\alpha$ and $2n$ wave functions. 
The spatial expansions, $B_c$, of the cluster distribution
from the origin and the cluster size, $b_c$,
are the same for each kind of clusters ($c=1$ for $\alpha$ and $c = 2, 3$ for $2n$), 
that is an $\alpha$ has $(B_{\alpha}, b_{\alpha})$ 
and both of dineutrons have $(B_{2n}, b_{2n})$. 
In the present $\alpha+2n+2n$ $\alpha$DC wave function, 
we take the $b_{\alpha}$ value as 1.4 fm which gives the minimum energy of a free $\alpha$. 
For the parameter $B_{\alpha}$, we choose as  
$B_{\alpha}^2 = ( \beta_{\alpha {\rm DC}}^2 - b_{\alpha}^2 ) / 2$ 
that satisfies the relation 
$2\beta_{\alpha {\rm DC}}^2 = 2B_{2n}^2 + 2b_{2n}^2 = 4B_{\alpha}^2 + 2b_{\alpha}^2$. 
Then, the total center of mass motion can be separated from the intrinsic wave function. 
Eq.~(\ref{eq:aDC_wf}) is rewritten after the $\boldsymbol{R}_c$- integral as
\begin{align}
& \Psi_{\alpha{\rm DC}}(B_{2n},b_{2n}) \nonumber \\ 
&\ \propto 
\mathcal{A} \{ \exp \Big[ - \frac{1}{\beta_{\alpha {\rm DC}}^2}
(4\boldsymbol{X}_1^2 + 2\boldsymbol{X}_2^2 + 2\boldsymbol{X}_3^2) \Big] \nonumber \\
&\ \hspace{2em} \times
\psi_{\alpha}(b_{\alpha}) \psi_{2n}(b_{2n}) \psi_{2n}(b_{2n}) \} \nonumber \\
&\ \propto \mathcal{A} \{ \exp \Big[ - \frac{1}{\beta_{\alpha{\rm DC}}^2}
(4\tilde{\boldsymbol{X}}_1^2
+ 2\tilde{\boldsymbol{X}}_2^2 + 2\tilde{\boldsymbol{X}}_3^2) \Big]  
\nonumber \\
&\ \hspace{2em} \times
\psi_{\alpha}(b_{\alpha}) \psi_{2n}(b_{2n}) \psi_{2n}(b_{2n}) \} \nonumber \\
&\ \hspace{4em} \times \psi_G(\beta_{\alpha{\rm DC}}), 
\label{eq:aDC_wf_int} \\
&\ \hspace{2em} \psi_G(\beta_{\alpha{\rm DC}})
\propto \exp \left[ - \frac{1}{\beta_{\alpha{\rm DC}}^2} 8 \boldsymbol{X}_G^2 \right], 
\end{align}
where $\boldsymbol{X}_c$ is the center of mass coordinate of each cluster
and $\boldsymbol{X}_G$ is the coordinate of the center of mass of the total system
($\boldsymbol{X}_G = \sum_c \boldsymbol{X}_c \ (c=1, 2, 3)$). 
$\tilde{\boldsymbol{X}}_c$ is defined as 
$\tilde{\boldsymbol{X}}_c \equiv \boldsymbol{X}_c - \boldsymbol{X}_G$. 
$\psi_{\alpha, 2n}$ is the intrinsic wave function of the $\alpha$ and the dineutron. 
Since we choose the value of $b_{\alpha}$ and $B_{\alpha}$ as mentioned above, 
the $\alpha$DC wave function actually contains two parameters, 
$b_{2n}$ (the dineutron size) and $B_{2n}$ (the dineutron expansion from the origin). 
As seen in Eq.~(\ref{eq:aDC_wf_int}), three clusters, that is one $\alpha$ and two dineutrons, 
are distributed around an origin in each $S$-orbit, 
whose expansion from the origin is characterized by $B_c$ or $\beta_{\alpha {\rm DC}}$, 
and the cluster size by $b_c$. 

The $\alpha$DC wave function contains only 
two independent parameters $b_{2n}$ and $B_{2n}$, 
and it can describe a condensation of two dineutrons 
in the same $S$-wave easily and clearly, 
as the $\alpha$ condensate wave function can. 
Although the form of the $\alpha$DC wave function (Eq.(\ref{eq:aDC_wf})) is derived from 
the $\alpha$ condensate wave function, 
the $\alpha$DC wave function has two main differences from the $\alpha$ condensate wave function; 
i) the $\alpha$DC wave function includes not only $\alpha$ clusters but also dineutron clusters, 
and ii) the size parameters, $b_c$, can be different for each kind of cluster, 
that is the dineutron size is different from the $\alpha$ size.  
The difference ii) is a crucial point to describe a system containing dineutron clusters, 
for a dineutron is much softer than an $\alpha$ 
(as discussed in Sec.~\ref{secII-C-3})
so that the assumption of the same size parameters for all of the $\alpha$ and dineutrons 
seems to be unreasonable. 
As in the DC wave function introduced in Sec.~\ref{secII}, 
we describe the system including dineutrons with various size and expansion
by changing $b_{2n}$ and $B_{2n}$ to various values. 
When $b_{2n}$ is small, 
the system contains two compact dineutrons and vice versa. 
In the case of large $B_{2n}$, two dineutrons occupy the $S$-orbit expanded largely, 
so the system with small $b_{2n}$ and large $B_{2n}$ is of our interest, 
the condensation of two compact dineutrons. 

In Sec.~\ref{secII}, 
we have suggested the existence of the dineutron condensation in the excited $0^+$ state of $^8$He,
and then we are interested in the formation mechanism of such a dineutron condensation. 
Since we omit the dineutron dissociation 
and assume that the system is composed of one $\alpha$ and two dineutrons
in the $S$-orbits around the origin 
in the present $\alpha+2n+2n$ $\alpha$DC wave function, 
its model space is a limited part of the one of the $^8$He DC wave function. 
However, the $\alpha+2n+2n$ $\alpha$DC wave function 
has the advantages that two dineutrons can be treated equally
and distributed in the same $S$-wave
and that the total center of mass motion can be removed exactly in a certain condition, 
analytically without any approximation. 
Therefore, the $\alpha$DC wave function is expected to be useful
to investigate the details of the dineutron condensation, 
though it is not suitable to study the ground state of $^8$He
where the breaking effect of the dineutrons becomes significant. 
With this framework, 
we analyze the dineutron behavior in the system of $\alpha+2n+2n$, 
that is the ideal system of $^8$He 
under the assumption of two spin-singlet pairs of two neutrons plus an $\alpha$. 
In order to consider the effect of the attraction from the core to the dineutron condensation, 
we also extend the $\alpha$DC wave function to the triton$(t)+2n+2n$ system, 
and compare the energetic behavior of the dineutrons in these systems. 
The formulation for the $t+2n+2n$ system is the same as that for the $\alpha+2n+2n$ system, 
just replacing an $\alpha$ by a $t$ composed of ($p\uparrow, n\uparrow, n\downarrow$). 
The $t+2n+2n$ system does not contain an $\alpha$
but we call even the one describing $t+2n+2n$
as the $\alpha$DC wave function for convenience. 
In the calculation of $t+2n+2n$, the size parameter of the $t$ is fixed to 1.59 fm
which gives the minimum energy of a free $t$, 
and the rest is not different from that in the $\alpha+2n+2n$ calculation. 

In this section, 
we calculate the expectation value of the Hamiltonian, 
$\langle \Psi_{\alpha {\rm DC}} (B_{2n}, b_{2n}) |H
| \Psi_{\alpha {\rm DC}} (B_{2n}, b_{2n}) \rangle$, 
as the function of the dineutron size, $b_{2n}$, 
and the expansion, $B_{2n}$, of the dineutron distribution around the origin. 
Then, we discuss the energy of the total system of $\alpha+2n+2n$ or $t+2n+2n$
depending on the dineutron behavior. 
For the Hamiltonian, $H$, 
we use the same one with the m55 parameter set
as the one in Sec.~\ref{secII}, 
except for the spin-orbit term. 
Here we omit the spin-orbit force in Eq.~(\ref{eq:hamiltonian_DC}), 
because, in the $\alpha+2n+2n$ system, 
each cluster is spin-singlet and the spin-orbit interaction does not have an effect. 
Also in the $t+2n+2n$ system, 
the cluster having non-zero spin is only one $t$ 
so that the effect of the spin-orbit interaction should be negligibly small. 
The aim of this section is to discuss the dineutron behavior in the dineutron condensation, 
so we adopt the m55 set as the parameter of the central force
which reproduces the unbound feature of a two-neutron pair. 
We have checked that the following discussion does not change qualitatively 
even if we use the m59 set where two neutrons are bound.

\subsection{Energy of two-dineutron condensation}
\label{secIII-B}

We discuss about the formation mechanism of the two-dineutron condensation 
around the $\alpha$ core. 
We apply the $\alpha+2n+2n$ $\alpha$DC wave function, 
and consider how the total energy of the dineutron condensate state 
depends on the size of dineutrons and the expansion of the dineutron distribution from the origin. 
In addition, we compare the energetic behavior in the $\alpha+2n+2n$ system 
with that in the $t+2n+2n$ system
to show that the moderate potential from the core 
is essential for the formation of the dineutron condensation.

\begin{figure}[h]
\includegraphics[scale=0.65]{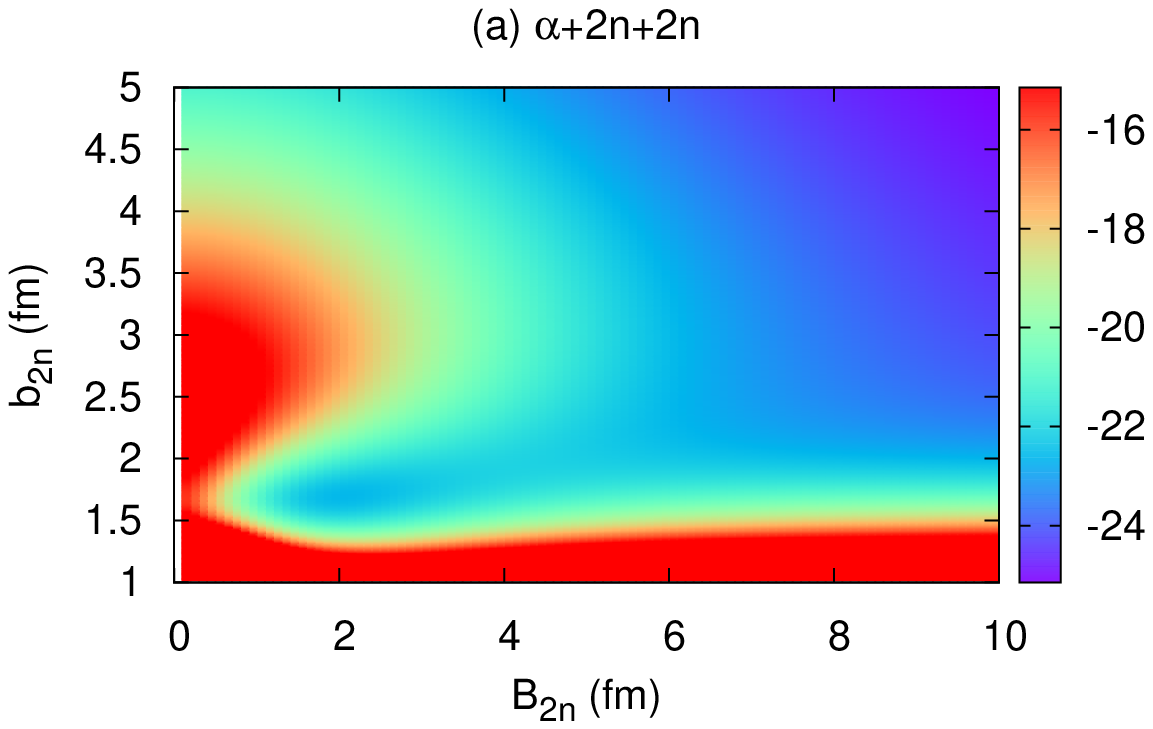} \\
\vspace{1em}
\includegraphics[scale=0.65]{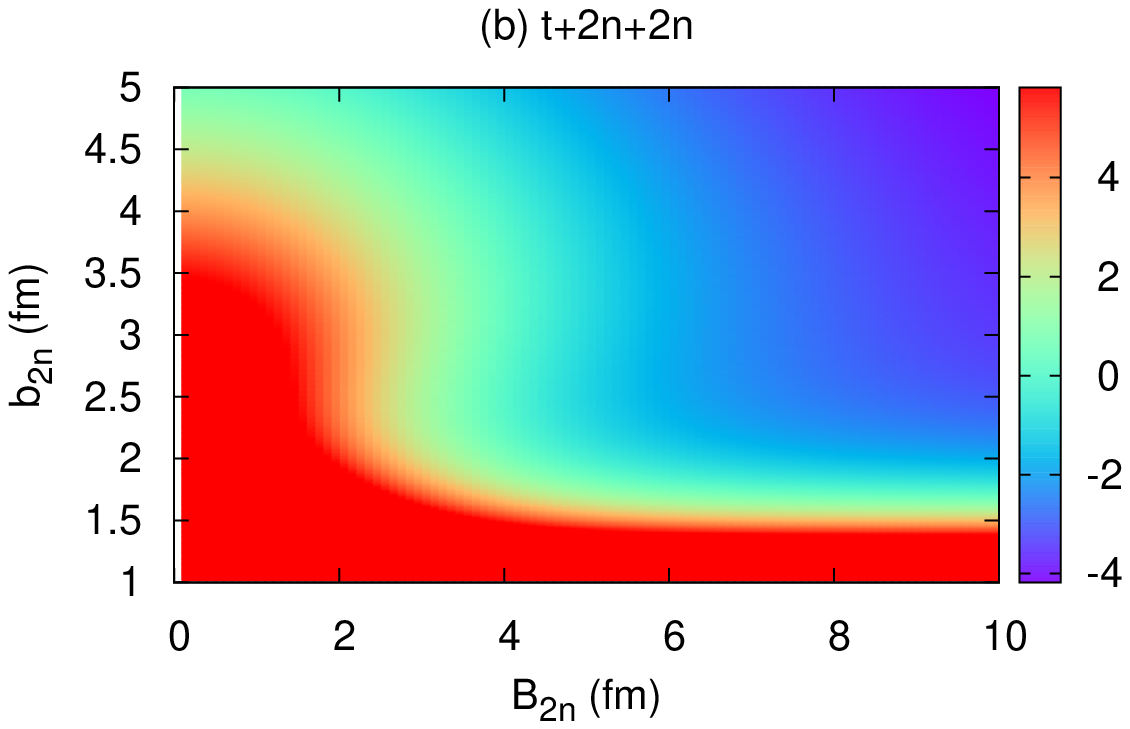} 
\caption{(Color online) The energy surface on the $B_{2n}$-$b_{2n}$ plane
calculated with the $\alpha+2n+2n$ and $t+2n+2n$ $\alpha$DC wave function
((a) and (b)).
Each of them are plotted by the energy region 
from the energy at $(B_{2n},b_{2n}) = (10.0,5.0)$ 
to that $+10$ MeV. }
\label{fig:Ene_surf_B_b_aDC}
\end{figure}
In order to discuss the dineutron behavior in the two-dineutron condensation around an $\alpha$, 
we show in Fig.~\ref{fig:Ene_surf_B_b_aDC}(a) the $B_{2n}$ and $b_{2n}$ dependence of 
the energy of the $\alpha+2n+2n$ $\alpha$DC wave function, 
which shows how the energy of the total system depends on the dineutron size and expansion. 
As seen in Fig.~\ref{fig:Ene_surf_B_b_aDC}(a), 
there appears an energy valley along the $b_{2n} \sim 1.7$ fm line 
in the region of $B_{2n} \lesssim 5$ fm, 
which corresponds to the states containing two compact dineutrons 
near and a little outside the nuclear surface.  
In addition, there exists a broad repulsive region around $(B_{2n},b_{2n}) \sim (0.0, 3.0)$
which is the states containing valence neutrons without the spatial correlation. 
It means that the $\alpha+2n+2n$ system favors energetically
the situation that the dineutron center of mass 
is expanded to some extent keeping its size compact.
The origin of these energy structures on the $B_{2n}$-$b_{2n}$ plane
is in principle the same as that of one dineutron around a core
discussed in Ref.~\cite{kobayashi11}:
The Pauli repulsive effect from the core contributes
to the repulsive region which prevents the dineutrons from spreading.  
This repulsive region and 
the competition between the kinetic and potential energy in each dineutron
results in the modestly compact dineutrons ($b_{2n} \sim 1.7$ fm). 

To begin with, we explain the origin
of the repulsive region around $(B_{2n},b_{2n}) \sim (0.0, 3.0)$. 
When $B_{2n} \sim 0$, 
each neutron in a dineutron is distributed in the $s$-wave around the origin, 
which corresponds to the configuration of two neutrons distributed spherically, 
in other word two spatially uncorrelated neutrons. 
Since the neutrons in the $\alpha$ core already occupy the $s$-orbit, 
a strong Pauli effect works between these $s$-orbit neutrons
and the energy in the region of $B_{2n} \sim 0$ is increased so largely. 
Therefore, the system containing an $\alpha$ core plus four valence neutrons 
without the spatial correlation is unfavorable energetically. 

Subsequently, we explain the origin 
of the energy valley along the $b_{2n} \sim 1.7$ fm line. 
In the small $b_{2n}$ region where a compact dineutron is formed, 
the kinetic energy of two neutrons in a dineutron is increased 
due to the uncertainty principle, 
but at the same time, they gain the attraction from each other. 
On the other hand, in the larger $b_{2n}$ region, 
both of the kinetic and potential energies are lowered. 
However, there is the repulsive area in the small $B_{2n}$ region
where the valence neutrons are distributed near the $\alpha$ core 
so that the pairs of two neutrons cannot be swelled near the $\alpha$ core, 
that is near the nuclear surface. 
Such competition of the kinetic and potential energies
and the repulsive effect explained above result in the energy valley on the small $b_{2n}$ line
in the relatively small $B_{2n}$ region. 
We conclude that, due to the repulsive region around the $B_{2n} \sim 0$ region, 
uncorrelated valence neutrons are strongly forbidden
and that, due to the energy valley along a small $b_{2n}$ line, 
two compact dineutrons distributed at the nuclear surface region are favored energetically. 
This valley is the key of the formation of the dineutron condensation. 
Such a formation mechanism of the dineutron condensation
is expected to be applicable to the other nuclei of some $\alpha$s and dineutrons. 

Next, we discuss the role of the core potential in the dineutron condensation
by comparing the energy behavior in $\alpha+2n+2n$ with that in $t+2n+2n$. 
There is only one crucial difference between these systems
that the attraction from a $t$ is much weaker than that from an $\alpha$
(seen in the difference of the absolute energies 
in Fig.~\ref{fig:Ene_surf_B_b_aDC}(a) and (b)), 
because of the smaller proton number. 
We plot the energy surfaces of $t+2n+2n$ on the $B_{2n}$-$b_{2n}$ plane
in Fig.~\ref{fig:Ene_surf_B_b_aDC}(b). 
At a glance, it is quite different from that of $\alpha+2n+2n$; 
there is no clear valley along the line of the certain small $b_{2n}$ value. 
As discussed in the $\alpha+2n+2n$ part, 
the energy valley along a small $b_{2n}$ line is the origin of the dineutron condensation. 
The fact that the valley disappears in the $t+2n+2n$ system
means that compact dineutrons are unlikely to be formed. 
It is because the interaction between a triton and valence neutrons 
is so weak that the neutrons are not attracted to a triton, 
unlike in the $\alpha+2n+2n$ system 
containing an $\alpha$ which draws valence neutrons closer. 
When the core attraction is too weak, 
two neutrons are expanded so largely that they behave as if in the free space. 
It is the point that the core attraction is strong enough to pull valence neutrons, 
but if it is too strong, the dineutron is dissociated to gain the spin-orbit energy 
as discussed in Sec.~\ref{secII}. 
Therefore, for the formation of the dineutron condensation, 
the attraction should be moderate. 
As suggested in the previous study of the dineutron correlation in the nuclear matter
\cite{matsuo06}, 
the matter density is the essential factor for the dineutron formation, 
and correspondingly in the finite nuclei, 
the attraction between the core and the valence neutrons
is the essence for the dineutron formation. 
The more close and realistic investigation is necessary to come to a conclusion
about the dineutron condensation in realistic $^7$H, 
as suggested in Ref.~\cite{aoyama09}, 
but we just remark that the moderate attraction from the core
is essential for the dineutron formation. 
It is difficult to describe the structure of $^7$H in detail since it is not a bound state, 
and it is a very challenging issue in future to investigate realistic $^7$H 
in order to make clear how the dineutrons behave in such an extreme situation. 

Let us consider the analogy with and the difference from the energy behavior of 
the $\alpha$ condensation in a $3\alpha$ system.
The dependences of the energy of the three-$\alpha$ condensation 
on the $\alpha$ cluster size, $b$, 
and the expansion of the $\alpha$ cluster distribution, $R_0$, were investigated 
by using the $\alpha$ condensate wave function 
in Ref.~\cite{tohsaki01}.  
In the energy surface of 
the three-$\alpha$ condensation on the $R_0$-$b$ plane (Fig.~1 in Ref.~\cite{tohsaki01}),
there is an energy pocket at the region of small $b$ and small $R_0$, 
which corresponds to the compact ground state of $^{12}$C. 
In addition, there is an energy valley along the constant $b$ line
and it is deep and extends clearly up to the large $R_0$ region, 
which is essentially the origin of the three-$\alpha$ condensate state of $^{12}$C$(0^+_2)$. 
Also in the case of the dineutron condensation in the ideal $\alpha+2n+2n$ system,
a local energy minimum ($(B_{2n}, b_{2n}) \sim (2.0, 1.6)$)
and an energy valley ($b_{2n} \sim 1.7$ fm) do exist on the $B_{2n}$-$b_{2n}$ plane 
(Fig.~\ref{fig:Ene_surf_B_b_aDC}(a)), 
but they are shallower than those in the three-$\alpha$ case 
and the valley disappears in the large $B_{2n}$ region 
where the attraction from the $\alpha$ core becomes weak. 
These behaviors reflect the fact that $\alpha$ clusters are rigid 
even in free space or in the deeply bound system, 
while dineutron clusters are fragile and their size can change largely 
depending on the condition, that is the nuclear density and so on.

\subsection{Relation between the realistic and ideal dineutron condensate systems}
\label{secIII-C}

Finally, we relate the above discussion of the ideal $\alpha+2n+2n$ system to 
the realistic $^8$He$(0^+)$ states and consider the feature of the dineutron condensation
in $^8$He($0^+_2$) suggested in Sec.~\ref{secII-C-3}. 
As mentioned before, $^8$He$(0^+_1)$ contains the $(0p_{3/2})^4$ shell-model 
component which is not included in the ideal $\alpha+2n+2n$ model space, and also 
some $\alpha+2n+2n$ component of two compact dineutrons at the very nuclear surface
(Figs.~\ref{fig:overlap_B_b_2DC_1st}). 
The $\alpha+2n+2n$ component in $^8$He$(0^+_1)$ 
corresponds to the tip of the energy valley in Fig.~\ref{fig:Ene_surf_B_b_aDC}(a). 
On the other hand,  
$^8$He$(0^+_2)$ contains the more developed and more swelled dineutron clusters.
The component of the dineutron condensation in the realistic $^8$He$(0^+_2)$ state
corresponds to the broad region of 
$B_{2n} \gtrsim 3$ fm and $1.5 \lesssim b_{2n} \lesssim 2.5$ fm along the valley 
in Fig.~\ref{fig:Ene_surf_B_b_aDC}(a). 
In such a region of relatively large $B_{2n}$ ($\gtrsim 3$ fm), 
the repulsive region is reduced so that a dineutron can be swelled to some extent. 
Then, we can reach the following understanding:
Under the orthogonal condition to $^8$He$(0^+_1)$ 
which partially contains the compact $\alpha+2n+2n$ state, 
the dineutron condensate state of $^8$He$(0^+_2)$ is constructed 
owing to the energy valley on the $B_{2n}$-$b_{2n}$ plane, 
in other words
because the energy changes softly toward the large $B_{2n}$ region 
in the $\alpha+2n+2n$ system containing compact dineutrons.

The origin of the dineutron condensate state of realistic $^8$He$(0^+_2)$ mentioned above 
has some similarity with that of the $\alpha$ condensate state of $^{12}$C$(0^+_2)$. 
However, there exist remarkable differences between the dineutron condensation in $^8$He 
and the $\alpha$ condensation in $^{12}$C,  
originating in the difference of the rigidity of constituent clusters, $\alpha$ or dineutron.
At first, as for the ground states, 
$^{12}$C$(0^+_1)$ is approximately described well within the model space 
of the $\alpha$ condensate wave function 
as the compact state of three $\alpha$ clusters near the origin \cite{tohsaki01}, 
while $^8$He$(0^+_1)$ contains not only the compact state of $\alpha+2n+2n$ 
but also the $(0p_{3/2})^4$ shell-model component 
out of the ideal $\alpha+2n+2n$ model space, 
because of the fragility of the dineutrons. 
Second, regarding the excited condensate states, 
$^{12}$C$(0^+_2)$, is described as the excited state orthogonal to the compact ground state
in the model space of the $\alpha$ condensate wave function. 
On the energy surface on the $R_0$-$b$ plane, 
this state corresponds to the three-$\alpha$ condensate wave function
at the large $R_0$ region on the deep energy valley along the constant $b$ line. 
On the other hand, 
$^8$He$(0^+_2)$, the candidate for the two-dineutron condensation, 
does not directly correspond to the $\alpha+2n+2n$ $\alpha$DC wave function 
on the energy valley along the constant $b_{2n}$ line 
in Fig.~\ref{fig:Ene_surf_B_b_aDC}(a), 
and this state mainly contains the component of larger dineutron clusters 
(Figs.~\ref{fig:overlap_B_b_2DC_2nd}). 
It is because, in the large $B_{2n}$ region, 
the repulsive area is reduced as shown in Fig.~\ref{fig:Ene_surf_B_b_aDC}(a) 
so that the dineutrons can be swelled. 
These differences originate in the softness and fragility of the dineutrons, 
and as a result, 
it is in general necessary to take into account the internal degrees of freedom, 
such as the breaking and size-changing effects, of the dineutrons explicitly, 
unlike rigid and stable $\alpha$ clusters.

\section{Summary}
\label{secIV}

In this work, 
we have investigated the dineutron correlation in the ground and excited states of $^8$He 
with the framework using an extended $^6$He$+2n$ cluster wave function and 
the dineutron condensate wave function. 
The former can describe the configurations of an $\alpha$ core 
and four valence neutrons in the $(0p)^4$-orbits 
and the ones of $^6$He in $(0p)^2$ and one dineutron around the $^6$He core.  
The latter can describe the configurations including 
one- and two-dineutron developed spatially far from the $^6$He and $\alpha$ core, respectively. 

In the ground state of $^8$He, 
a part of dineutrons is dissociated due to the spin-orbit force because of their moderate binding, 
that is to say the shell-model configuration and the one-dineutron configuration are mixed
due to the competition between the $j$-$j$ coupling feature and the dineutron correlation. 
On the other hand, the component of two dineutrons
distributed far from the $\alpha$ core is minor in the ground state
because of the fragility of the dineutrons, 
and instead, it contributes to the $0^+_2$ state mentioned below.
In addition, we compare the dineutron component in the ground state of $^8$He
with that in the ground state of $^6$He. 
In the ground state of $^6$He which is an extremely loosely bound system, 
the one-dineutron configuration is dominant, 
while the shell-model configuration is minor because of the weak attraction
between an $\alpha$ and two valence neutrons. 
As a conclusion, we consider that the strength of the binding from the core
is a key for the formation and breaking of a dineutron. 

We have also suggested the excited $0^+$ state in $^8$He
which includes a dineutron condensation. 
This state is characterized by a gas-like structure of 
largely developed one $\alpha$ and two dineutron clusters. 
The dineutron size is much larger than that at the nuclear surface in the ground state. 
It is an interesting feature of the dineutron correlation
that the strength of the correlation changes significantly 
depending on the distance from the core and the number of dineutrons. 
This feature may be associated with the density dependence 
of the dineutron correlation in nuclear matter. 
Such an expanded structure results in the possibility 
that this condensate state would be confirmed by measuring the monopole transition to the ground state. 

We have discussed the formation mechanism of the dineutron condensation in $^8$He 
by a new kind of condensate wave function, 
the $\alpha$ and dineutron condensation, 
and investigated the energy behavior of two dineutrons in the $\alpha+2n+2n$ system. 
We made it clear that two compact dineutron can be formed around an $\alpha$
and can be distributed far from the $\alpha$ somewhat, 
and such a behavior is the origin of the two-dineutron condensation in $^8$He. 
We have done the same analysis for the triton$+2n+2n$ system 
and found that the energy behavior of two dineutrons around a triton changes drastically 
from that around an $\alpha$,
and two compact dineutrons are formed no longer. 
Such a drastic difference in the dineutron behavior around an $\alpha$ and around a triton 
suggests that the attraction from a core 
plays an important role for the formation mechanism of the dineutron condensation. 

In future, 
we would like to investigate the more neutron-rich nuclei such as $^7$H and $^{10}$He
from the viewpoint of the dineutron condensation, 
and make clear the dineutron behavior in such an extreme situation. 
In addition, we are also interested in the condensation of some $\alpha$s and dineutrons
in neutron-rich nuclei such as $^{14}$C (the Hoyle state plus one dineutron)
and heavier C isotopes. 
The method of the present $\alpha$DC wave function can be easily extended 
to general systems having some $\alpha$s and some dineutrons. 
It is interesting problem whether condensate state containing 
two species of bosons, $\alpha$s and dineutrons, exist in excited states of neutron-rich nuclei.

\appendix

\section{Approximation of two-dineutron condensate wave function}
\label{appendix1}

In this appendix, we explain the two-dineutron condensate (2DC) wave function, 
$\Phi_{2 {\rm DC}}(\beta, b_n)$, 
in the calculation of the overlap, 
$\mathcal{N}_{2{\rm DC}}(\beta,b_n)$ of Eq.~(\ref{eq:overlap_2DC}), 
which is used for the analysis of the component of the two-dineutron condensation. 

The two-dineutron condensation is the state with two dineutrons in the $S$-wave
around a core. 
If we neglect the recoil effect of the core with respect to dineutrons, 
the 2DC wave function of the $\alpha+2n+2n$ system can be written as
\begin{align}
\Phi_{2 {\rm DC}}^{\rm norecoil} 
= &\ \frac{1}{\sqrt{8!}}\int d^3 \boldsymbol{R}_{2n_1} \ d^3 \boldsymbol{R}_{2n_2}
\ \exp \left[ - \frac{\boldsymbol{R}_{2n_1}^2}{B_{2n_1}^2} 
- \frac{\boldsymbol{R}_{2n_2}^2}{B_{2n_2}^2} \right] 
\nonumber \\
&\ \times
\mathcal{A} \{ \Phi_{\alpha}(\boldsymbol{R}_{\alpha}=0, b_{\alpha}) \nonumber \\
&\ \hspace{2em} \times
\Phi_{2n}(\boldsymbol{R}_{2n_1}, b_{2n}) \Phi_{2n}(\boldsymbol{R}_{2n_2}, b_{2n}) 
\} \nonumber \\
= &\ \frac{1}{\sqrt{8!}}\int d^3 \boldsymbol{R}_{2n_1} \ d^3 \boldsymbol{R}_{2n_2}
\ \exp \left[ - \frac{\boldsymbol{R}_{2n_1}^2}{B_{2n_1}^2} 
- \frac{\boldsymbol{R}_{2n_2}^2}{B_{2n_2}^2} \right] 
\nonumber \\
&\ \times \mathcal{A} \Big\{ 
\exp \Big[ 
- \frac{4}{2b_{\alpha}^2} \boldsymbol{X}_{\alpha}^2 
- \frac{2}{2b_{2n}^2} (\boldsymbol{X}_{2n_1} - \boldsymbol{R}_{2n_1})^2 \nonumber \\
&\ \hspace{5em}
- \frac{2}{2b_{2n}^2} (\boldsymbol{X}_{2n_2} - \boldsymbol{R}_{2n_2})^2 \Big] \nonumber \\
&\ \hspace{2em}
\times \psi_{\alpha} \psi_{2n} \psi_{2n} \Big\}, \label{eq:exact_2DCwf}
\end{align}
where $\boldsymbol{R}_c$ is the Gaussian center 
of the single particle wave functions composing the cluster of $c = (\alpha, 2n_1, 2n_2)$. 
(Here we label two dineutrons as $2n_1$ and $2n_2$.)
$\boldsymbol{X}_c$ is the center of mass coordinate of the cluster $c$, 
$\boldsymbol{X}_c = \sum_{i \in c} \boldsymbol{r}_i$. 
$\psi_c$ is the internal wave function of the cluster $c$. 
In the heavy mass limit of the core, 
the wave function $\Phi_{2{\rm DC}}^{\rm norecoil}$
with $B_{2n_1} = B_{2n_2}$ exactly describes 
the ideal two-dineutron condensation 
where two identical dineutrons with the same size, $b_{2n}$, occupy the same $S$-orbit
whose expansion is characterized by $B_{2n_1} (= B_{2n_2})$. 
For the $\alpha$-cluster core, we should consider the recoil effect. 
However, it is difficult to treat the recoil effect and remove the center of mass motion exactly 
in the present framework for $^8$He. 
Thus we prepare the approximated 2DC wave function, 
which hereafter we simply call the 2DC wave function, $\Phi_{2{\rm DC}}$, 
and use it to evaluate the two-dineutron component in the $^8$He wave function. 
The 2DC wave function is constructed by the superposition of DC wave functions
having $\alpha+2n$ plus a dineutron in the $S$-wave. 
For consistency with the DC wave functions used in the $^8$He calculation, 
we adopt a choice of $\boldsymbol{R}_{\alpha} = - \boldsymbol{R}_{2n_2}/2$. 
This treatment means that the recoil effect from only one-dineutron is taken into account. 
By performing the integrals with respect to 
$\boldsymbol{R}_{2n_1}$ and $\boldsymbol{R}_{2n_2}$ analytically, 
Eq.~(\ref{eq:exact_2DCwf}) is rewritten as 
\begin{align}
\Phi_{\rm 2DC} = &\ \frac{1}{\sqrt{8!}}
\mathcal{A} \Big\{ \exp \Big[ - \frac{1}{\beta_{2n_1}^2} X_{2n_1}^2 \nonumber \\
& \hspace{6em}
- \frac{b_{\alpha}^2}{\beta_{2n_2}^2 b_{\alpha}^2 
+ 1/2 \ B_{2n_2}^2 b_{2n}^2} X_{2n_2}^2 \nonumber \\ 
& \hspace{6em} 
- \frac{2b_{2n_2}^2}{\beta_{2n_2}^2 b_{\alpha}^2 
+ 1/2 \ B_{2n_2}^2 b_{2n}^2} X_{\alpha}^2 \nonumber \\
& \hspace{6em}
+ \frac{9/2 \ B_{2n_2}^2}{\beta_{2n_2}^2 b_{\alpha}^2 
+ 1/2 \ B_{2n_2}^2 b_{2n}^2} \delta X^2 \Big] \nonumber \\
&\ \hspace{3em} \times \psi_{\alpha} \psi_{2n} \psi_{2n}
\Big\}, \label{eq:2DC_wf}
\end{align}
where $\beta_{2n_k} \ (k = 1,2)$ is defined as $\beta_{2n_k}^2 = B_{2n_k}^2 + b_{2n}^2$.  
$\delta X$ is the $\alpha+2n$ center of mass coordinate 
defined as $\delta \boldsymbol{X} = 
(4\boldsymbol{X}_{\alpha} + 2\boldsymbol{X}_{2n_2})/6 $. 
In order to describe the two-dineutron condensation, 
we take $B_{2n_2}$ as 
$1/B_{2n_2}^2 = 1/B_{2n_1}^2 \times ( 1 + b_{2n}^2/2b_{\alpha}^2)$. 
Finally, if we neglect the term of $\delta X^2$ in Eq.~(\ref{eq:2DC_wf}), 
the 2DC wave function can be written as follows. 
\begin{align}
\Phi_{\rm 2DC} \simeq &\ \frac{1}{\sqrt{8!}}
\mathcal{A} \Big\{ \exp \Big[ - \frac{1}{\beta_{2n_1}^2}  \left( X_{2n_1}^2 + X_{2n_2}^2 
+ \frac{2b_{2n}^2}{b_{\alpha}^2} X_{\alpha}^2 \right) \Big] \nonumber \\
&\ \hspace{3em} \times \psi_{\alpha} \psi_{2n} \psi_{2n} \Big\}, 
\label{eq:2DC_wf_2}
\end{align}
where $\beta = \beta_{2n_1}$.
The right-hand equation indicates that
two same-size dineutrons are distributed around the origin in the $S$-orbit
whose expansion is characterized by $\beta$, 
so a dineutron condensate state can be described approximately. 
We comment that, when the size of the $\alpha$ and the dineutrons is the same, 
the ratio of the factor in front of $X_{\alpha}^2$ and that of $X_{2n_1,2n_2}$ is 4:2
so that the total center of mass motion can be extracted. 

Practically, while the $\boldsymbol{R}_{2n_1}$-integral is performed analytically, 
the $\boldsymbol{R}_{2n_2}$-integral is done numerically. 
We replace the $\boldsymbol{R}_{2n_2}$-integral
with the angular integral and the summation of the discretized coordinates
at regular intervals, $\Delta R$, 
that is $d^3 \boldsymbol{R}_{2n_2} 
\rightarrow d\boldsymbol{\Omega}_R \times \sum_i R_i^2 \Delta R$. 
In the present calculation, 
we choose from $R_1 = 2/3$ fm to $R_{30} = 20$ fm with $\Delta R = 2/3$ fm. 

In the investigation in Sec.~\ref{secII}, 
we use the approximation and the numerical integral mentioned above
to calculate the overlap with the 2DC wave function, $\Phi_{2{\rm DC}}(\beta, b_n)$ 
in Eq.~(\ref{eq:2DC_wf_2}), 
shown in Figs.~\ref{fig:overlap_B_b_2DC_1st} and \ref{fig:overlap_B_b_2DC_2nd}.

\section{Identification of $^8$He$(0^+_2)$; pseudo potential method}
\label{appendix2}

\begin{figure}[h]
\includegraphics[scale=0.65]{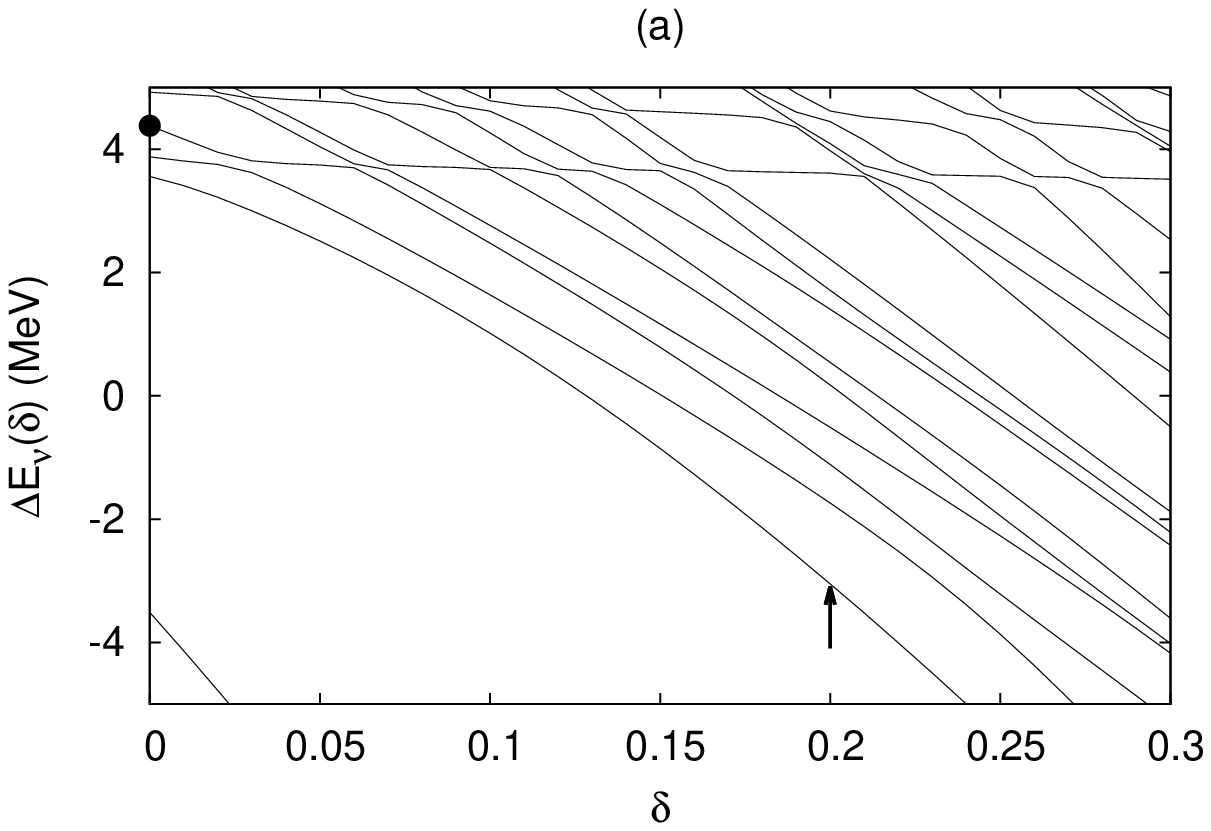} \\
\vspace{1em}
\includegraphics[scale=0.65]{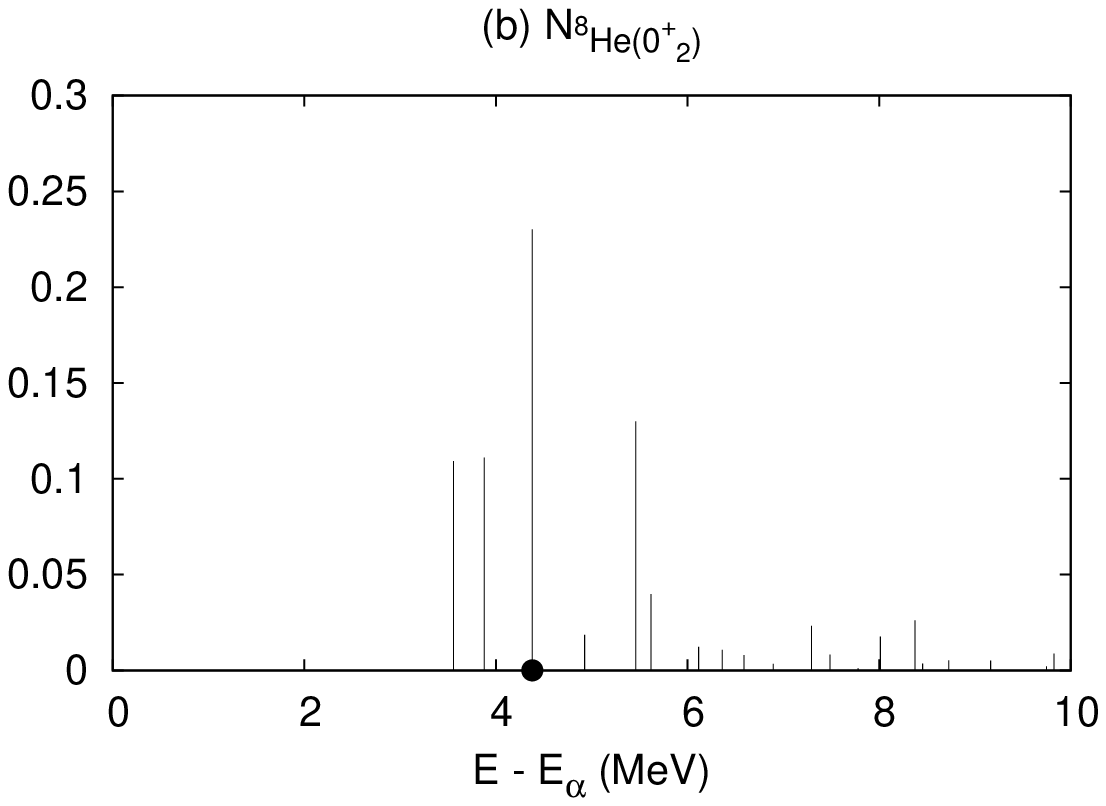} 
\caption{(a)The change of the energies of the $0^+$ states of $^8$He 
when the strength of the pseudo potential, $\delta$, changes, 
calculated with the m55 set.
The $^8$He$(0^+_2)$ state is marked with the dot at $\delta = 0$.
The arrow indicates the state used as the overlapped state in (b). \\
(b)The overlaps of the $0^+$ states of $^8$He
with the state indicated by the arrow in (a). 
The horizontal axis is the energy difference of each state and an $\alpha$ 
calculated with the original Hamiltonian. 
The dot indicates $^8$He$(0^+_2)$.}
\label{fig:E_overlap_ACCC}
\end{figure}
In the present results of $^8$He, 
$^8$He$(0^+_2)$ is the resonance state embedded in the continuum states 
such as $\alpha+4n$. 
Because of the largely expanded structure, 
it is likely to strongly couple to the continuum states. 
In order to identify such a resonance state simply, 
we use the pseudo potential method introduced in Refs.~\cite{enyo12, kobayashi12}. 
This method is based on the concept 
that physical resonance states gain the energy to be bound states
if the sufficient attraction is imposed 
and, as a result, they can be extracted from neighboring continuum states.
In this method, we add the artificial attraction by hand
to the original Hamiltonian, $H$ (Eq.~(\ref{eq:hamiltonian_DC})), as
\begin{gather}
\tilde{H}(\delta) = H + \delta \times \tilde{V}, 
\label{eq:pseudo_pot} \\
\tilde{V} = \sum_{i<j} v_0 \exp \left[ - \frac{r_{ij}^2}{a_0^2} \right], 
\end{gather}
where $\tilde{V}$ is the Gaussian-type short-ragne pseudo potential. 
Its parameters are fixed as $v_0 = - 100$ MeV and $a_0 = 1.0$ fm. 
The parameter $\delta$ controls the strength of the pseudo potential. 
By diagonalizing the Hamiltonian matrix for $\tilde{H}(\delta)$ 
with respect to the bases of Eq.~(\ref{eq:He8_wf}), 
we obtain energy levels of the $0^+$ states
as a function of the strength $\delta$ of the pseudo potential. 

As an example, 
we show the results in the m55 case. 
We subtract the energy of an $\alpha$ calculated with this Hamiltonian (Eq.~(\ref{eq:pseudo_pot}))
from the energies of the obtained $0^+_{\nu}$ states for each $\delta$ value. 
\begin{equation}
\Delta E_{\nu} (\delta) \equiv \langle \Psi_{^8{\rm He}(0^+_{\nu})}(\delta)| \tilde{H}(\delta) 
|\Psi_{^8{\rm He}(0^+_{\nu})}(\delta) \rangle
- \langle \Phi_{\alpha}| \tilde{H}(\delta) |\Phi_{\alpha} \rangle, 
\end{equation}
which are plotted in Fig.~\ref{fig:E_overlap_ACCC}(a). 
At $\delta = 0$, 
there exists a bound state at $\Delta E \sim -3.5$ MeV
which corresponds to the ground state. 
Increasing the strength $\delta$ of the pseudo potential, 
the energies of the $\alpha+4n$ continuum states draw plateau lines, 
while the resonance state comes down below the $\alpha+4n$ threshold. 
There are also many continuum states in other channels than $\alpha+4n$
such as $^6$He$+2n$ and $\alpha+2n+2n$. 
These continuum states come down to lower than the $\alpha+4n$ threshold
with the increase of $\delta$.  
Nevertheless, when the pseudo potential is sufficiently strong, 
the resonance state that is confined to finite space 
gains more energy and its energy position is lowered than those continuum states. 
At $\delta = 0.2$, 
we can identify $^8$He$(0^+_2)$ at $\Delta E_{\nu}(\delta = 0.2) \approx - 3.0$ MeV
as a pseudo-bound state decoupled from continuum states
(the first line from the bottom at $\delta = 0.2$
indicated with an arrow in Fig.~\ref{fig:E_overlap_ACCC}(a)). 
In order to see how much component of the resonance state 
the states calculated with the original Hamiltonian
(Eq.~(\ref{eq:hamiltonian_DC})) contain, 
we calculate their overlaps 
with the pseudo-bound state obtained with $\delta = 0.2$. 
The calculated overlaps are plotted in Fig.~\ref{fig:E_overlap_ACCC}(b). 
The amplitude of the pseudo-bound state is fragmented to some states
because of the coupling with continuum states.  
The $0^+$ state at $E-E_{\alpha} = 4.3$ MeV
marked with a dot in Fig.~\ref{fig:E_overlap_ACCC}(a) and (b)
has the most significant amplitude, 
indicating that it contains the main component of the resonance state.
Combining this result and the one of the significant two-dineutron component 
shown in Sec.~\ref{secII-C-3}, 
we identify the $0^+$ state ($E-E_{\alpha}=4.3$ MeV) as $^8$He$(0^+_2)$ of our interest. 
Also in the m59 case, 
we perform the same analysis to identify $^8$He$(0^+_2)$.

\begin{acknowledgments}
This work was supported by Grant-in-Aid for Scientific Research 
from Japan Society for the Promotion of Science (JSPS).
It was also supported by
the Grant-in-Aid for the Global COE Program ``The Next Generation of Physics,
Spun from Universality and Emergence'' 
from the Ministry of Education, Culture, Sports, Science and Technology (MEXT) of Japan.
A part of the computational calculations of this work was performed by using the
supercomputers at YITP.
\end{acknowledgments}

\bibliography{reference}

\end{document}